\documentclass{article}
\usepackage{ijcai22}

\usepackage{times}
\usepackage{soul}
\usepackage{url}
\usepackage[hidelinks]{hyperref}
\usepackage[utf8]{inputenc}
\usepackage[small]{caption}
\usepackage{graphicx}
\usepackage{amsmath}
\usepackage{amsthm}
\usepackage{booktabs}
\usepackage{algorithm}
\usepackage{algorithmic}
\usepackage{multirow}                %multirow for format of table
\usepackage{makecell}
\usepackage{color}
\usepackage{amsmath,amssymb,amsfonts}
\usepackage{array}
\usepackage{multirow}                %multirow for format of table
\usepackage{bm}
\def\1{\mathbf{1}}
\def\0{\mathbf{0}}

\def\A{\bm{A}}

\def\K{\bm{K}}

\def\I{\bm{I}}
\def\M{\bm{M}}

\def\X{\bm{X}}
\def\Y{\bm{Y}}

\def\mM{\mathcal{M}}

\def\mC{\mathcal{C}}

\def\mK{\bm{K}}

\def\mD{\mathcal{D}}

\def\mP{\mathcal{P}}

\title{Adaptive Convolutional Dictionary Network for CT Metal Artifact Reduction}

\author{
Hong Wang$^{1,2}$
\and
Yuexiang Li$^{1,}$\footnotemark[2]\and
Deyu Meng$^{2,3,}$\footnotemark[2]\and
Yefeng Zheng$^1$
\affiliations
$^1$Tencent Jarvis Lab, Shenzhen, China \\
$^2$Xi’an Jiaotong University, Xi'an, China \\
$^3$Peng Cheng Laboratory, Shenzhen, China\\
\emails
\{hazelhwang, vicyxli, yefengzheng\}@tencent.com, dymeng@mail.xjtu.edu.cn
}
\begin{document}
\maketitle
\renewcommand{\thefootnote}{\fnsymbol{footnote}}
\footnotetext[2]{Corresponding author}
\begin{abstract}
Inspired by the great success of deep neural networks, learning-based methods have gained promising performances for metal artifact reduction (MAR) in computed tomography (CT) images. However, most of the existing approaches put less emphasis on modelling and embedding the intrinsic prior knowledge underlying this specific MAR task into their network designs. Against this issue, we propose an adaptive convolutional dictionary network (ACDNet), which leverages both model-based and learning-based methods. Specifically, we explore the prior structures of metal artifacts, \emph{e.g.}, non-local repetitive streaking patterns, and encode them as an explicit weighted convolutional dictionary model. Then, a simple-yet-effective algorithm is carefully designed to solve the model. By unfolding every iterative substep of the proposed algorithm into a network module, we explicitly embed the prior structure into a deep network, \emph{i.e.,} a clear interpretability for the MAR task. Furthermore, our ACDNet 
can automatically learn the prior for artifact-free CT images via training data and adaptively adjust the representation kernels for each input CT image based on its content.
Hence, our method inherits the clear interpretability of model-based methods and maintains the powerful representation ability of learning-based methods. Comprehensive experiments executed on synthetic and clinical datasets show the superiority of our ACDNet in terms of effectiveness and model generalization. {\color{blue}{{\textit{Code is available at {\url{https://github.com/hongwang01/ACDNet}}.}}}}

\end{abstract}

% It's well-known that 
% during the CT imaging process,
\section{Introduction}
X-ray computed tomography (CT) has been broadly adopted for clinical diagnosis. Nevertheless, common metallic implants within patients, such as dental fillings and hip prosthesis, would adversely cause the missing of projection data during CT imaging, and thus lead to the obvious streaking artifacts and shadings in the reconstructed CT images. A robust model, which automatically reduces the unsatisfactory metal artifacts and improves the quality of CT images for subsequent clinical treatment, is worthwhile to develop~\cite{wang2021indudonet,wang2021dicdnet,lin2019dudonet}.

Against this metal artifact reduction (MAR) task, traditional model-based methods focused on reconstructing artifact-reduced CT images by filling the metal-affected region in sinogram with different estimation strategies, such as linear interpolation (LI)~\cite{kalender1987reduction} and normalized MAR~\cite{meyer2010normalized}. However, these methods always bring secondary artifacts in the restored CT images, since the estimated sinogram cannot finely meet the CT physical imaging constraint. There is another type of conventional MAR methods, which iteratively reconstruct clean CT images from unaffected sinogram~\cite{zhang2011metal} or weighted/corrected sinogram~\cite{lemmens2008suppression}.
Although such model-based methods are usually algorithmically interpretable, the pre-designed regularizers cannot flexibly represent the complicated artifacts in different metal-corrupted CT images collected from real applications. 
\begin{figure}[t]
  \begin{center}
     \includegraphics[width=1\linewidth]{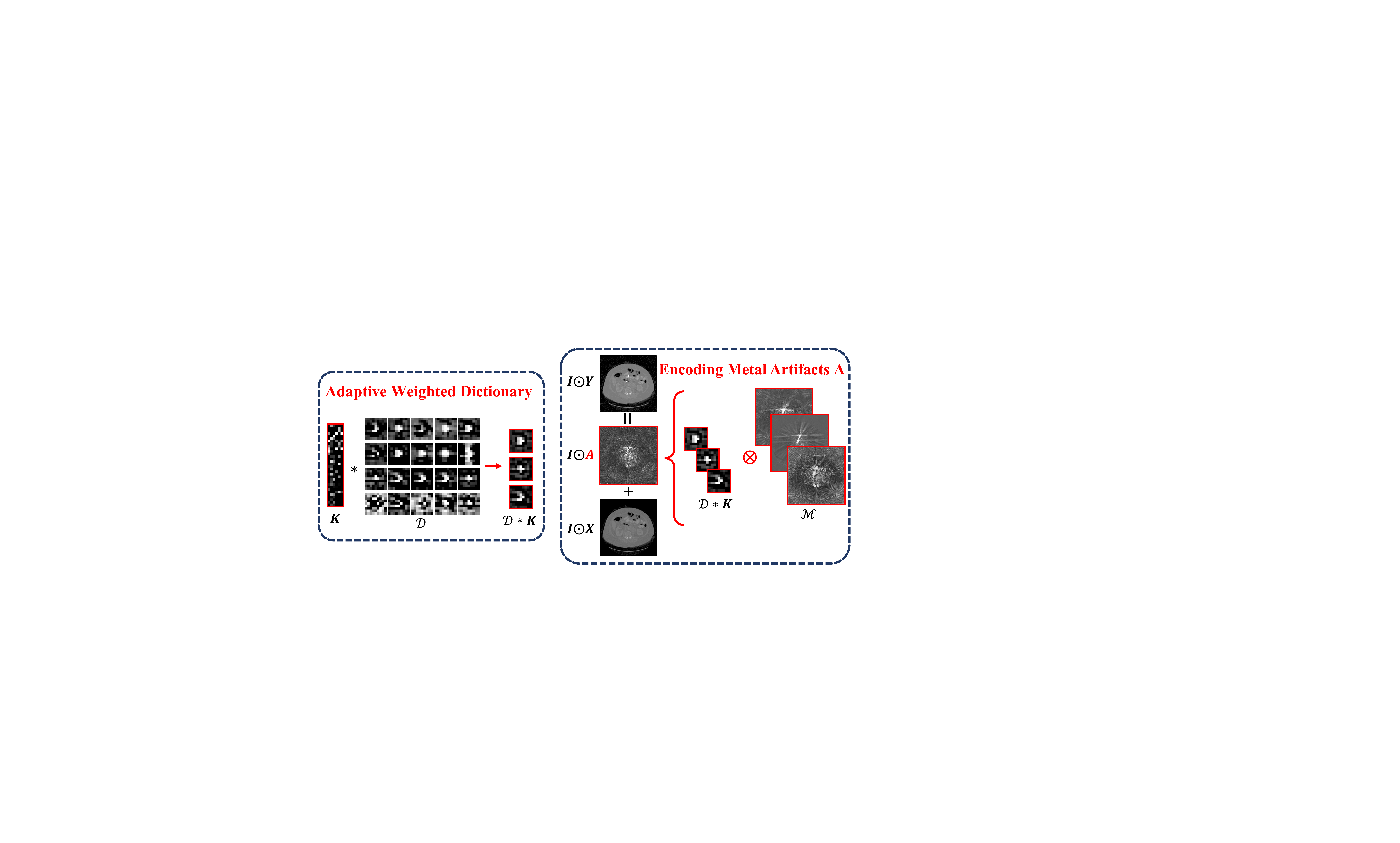}
  \end{center}
  \vspace{-4mm}
     \caption{The proposed weighted convolutional dictionary model for encoding metal artifacts $\A$ as $\left(\mD\ast\mK\right)\otimes\mM$. Here the dictionary $\mD$ is sample-invariant and the weighting coefficient $\K$ is sample-variant. By updating $\K$, the representation kernel for artifacts (\emph{i.e.}, $\mD\ast\mK$) can be adaptively inferred for every input image $\Y$.}
  \label{figintro}
    \vspace{-4mm}
\end{figure}

With the rapid development of deep learning (DL), recent years have witnessed the promising progress of DL for the MAR task. Some early works adopted convolutional neural network (CNN) to reconstruct clean sinograms and then utilized the filtered-back-projection transformation to reconstruct the artifact-reduced CT images~\cite{zhang2018convolutional,ghani2019fast}. Later, researchers adopted different learning strategies, \emph{e.g.,} residual learning~\cite{huang2018metal} and adversarial learning~\cite{wang2018conditional}, to directly learn the artifact-reduced CT images. Very recently, there is a new research line for the MAR task, which focuses on the mutual learning of sinograms and CT images~\cite{lin2019dudonet,lyu2020dudonet++,wang2021indudonet,wang2021indudonet+}.

Attributed to the powerful learning ability of CNN, deep MAR methods generally achieve superior performance over conventional model-based approaches. Yet, there is still some room for further performance improvement. First, most of current deep MAR works pay less attention to exploring the intrinsic prior knowledge of the specific MAR task, such as non-local streaking structures of metal artifacts (see Fig.~\ref{figintro}). The explicit embedding of such prior information can assist to regularize the solution space for metal artifact extraction, and further boost the MAR performance of deep networks as well as its generalization ability. Second, the involved image enhancement modules in current deep MAR works~\cite{lin2019dudonet,lyu2020dudonet++} are roughly the variants of U-Net. With such a general network design, the physical interpretability of every network module is unclear.

In this paper, we propose an explicit model to encode the prior observations underlying the MAR task and fully embed it into an adaptive convolutional dictionary network (namely ACDNet). The proposed framework inherits the clear interpretability of model-based methods and maintains the powerful representation ability of learning-based methods. In summary, our main contributions can be concluded as:

% To deal with the aforementioned problems, in this paper, we propose an approach to explicitly embed the prior observations underlying the MAR tasks into the deep learning framework, which accordingly increases the interpretability.

% \vspace{1mm}
\textbf{1) Prior Formulation.}
% \paragraph{1) Adaptive Prior Formulation.}
 For the MAR task, we explore that artifacts in different metal-corrupted CT images always present \textbf{common (\emph{i.e.}, sample-invariant)} patterns, such as non-local repetitive streaking structures, but with the \textbf{specific (\emph{i.e.}, sample-invariant)} intensity variation in CT images. Based on such prior observations, we adaptively encode the artifacts in every metal-corrupted CT image as a weighted convolutional dictionary (WCD) model (see Fig.~\ref{figintro}). 

% \vspace{1mm}
\textbf{2) Prior Embedding and Interpretability.}
% \paragraph{2) Prior Embedding and Clear Interpretability.}
To solve the WCD model, we propose an iterative algorithm with only simple operators and easily construct the ACDNet by unfolding every iterative step into the corresponding network module. Similar to learning-based methods, ACDNet automatically learns the priors for artifact-free CT images in a purely data-driven manner, overcoming the disadvantages of hand-crafted priors. Similar to model-based methods, ACDNet is explicitly integrated with the WCD model of artifacts and has clear physical interpretability corresponding to our optimization algorithm. The proposed ACDNet thus integrates the advantages of model-based and learning-based methodologies.

%  has clear physical interpretability corresponding to our optimization algorithm. These factors facilitate general users to understand the mechanism of every network connection.

% \vspace{1mm}
\textbf{3) Fine Generalization.} 
% \paragraph{3) Fine Generalizability.} 
With the regularization of the explicit WCD model, ACDNet can more accurately extract artifacts complying with prior structures. Comprehensive experiments, including cross-body-site generalization and synthesis-to-clinical generalization, finely substantiate the superiority of our method as well as excellent computation efficiency. Especially, as an image-domain-based method without utilizing sinogram, our ACDNet can be adopted as a plug-in module, which is easily integrated into current computer-aided diagnosis systems to deal with some practical scenarios where the projection data is difficult to acquire.

\section{Methodology}\label{sec:model}
In this section, we propose an explicit model to adaptively encode the priors of every metal-affected CT image and derive a simple algorithm for the subsequent network design.

% carefully analyze the prior characteristics of metal-affected CT images and propose an explicit model for encode such observations. establish a maximum-a-posterior-based optimization model for the MAR task. Then, based on a derived simple algorithm

\subsection{Weighted Convolutional Dictionary Model}
For an observed metal-corrupted CT image $\Y$, it is composed of two regions, \emph{i.e.}, metal part and non-metal part. Since metals generally have higher CT values than normal tissues, we ignore the information in the metal region and make efforts to reconstruct the non-metal region of $\Y$. Therefore, the decomposition model can be derived as:
\begin{equation}\label{e1}
\I\odot\Y=\I\odot\X +\I\odot\A,
\end{equation}
where $\I$ is a binary non-metal mask; $\X\in\mathbb{R}^{H\times W}$ is the to-be-estimated clean CT image; $\A$ is the to-be-extracted metal artifacts; $\odot$ is an element-wise multiplication.

From Eq.~\eqref{e1}, the estimation of $\X$ and $\A$ from $\Y$ is an ill-posed inverse problem. Against this issue, most of current learning-based methods empirically design complicated network architectures to directly learn $\X$ from $\Y$, which lack the consideration of \textbf{explicit prior knowledge} of MAR (\emph{e.g.,} the patterns of metal artifacts). However, embedding the explicit prior is helpful to finely regularize the solution space of such an ill-posed problem, and thus further improves the performance of deep networks, especially generalization ability.
% It's easy to understand that the embedding of \textbf{explicit prior knowledge} would be helpful to finely regularize the solution space of such an ill-posed problem and thus further improve the performance of deep networks, especially generalization ability. 
Based on this motivation, we explore the specific prior structures underlying the MAR task, and then propose a strategy to explicitly embed them into deep networks.

\paragraph{Prior Formulation.} Specifically, we disclose that for different metal-affected CT images, the metal artifacts are with roughly \textbf{common} patterns, \emph{e.g.}, non-local repetitive streaking structures. Besides, due to the mutual influence between normal tissues and metal artifacts, the patterns of metal artifacts in different CT images are not exactly the same and generally have some \textbf{specific} characteristics, \emph{e.g.}, pixel intensities of metal artifacts vary between different metal-corrupted CT images. Based on these prior observations, we formulate a weighted convolutional dictionary (WCD) model to encode the metal artifacts $\A$ as~\cite{wang2021rcdnet}:
%\footnote{Consistent to current SOTA works~\cite{lin2019dudonet, yu2020deep, lyu2020dudonet++}, the non-metal mask $\I$ as well as the metal mask $(\bm{1}-\I)$ is known, which is detailed in Sec.~\ref{sec:exp}.}
% Particularly, as shown in Fig.~\ref{figcd}, the metal artifacts repeatedly occur in different locations of metal-corrupted CT images with similar local streaking and star-shape patterns~\cite{lin2019dudonet,lyu2020dudonet++}. {\color{black}{Inspired by the work~\cite{wang2020model}}}, we propose a convolutional dictionary mechanism to encode the structural prior knowledge of $\A$:
\begin{equation}\label{A}
    \A = \sum_{n=1}^{N} \left(\mD\ast \K_{n} \right)\otimes \M_{n} = \left(\mD\ast\mK\right)\otimes\mM,
\end{equation}
where $\mD\in\mathbb{R}^{p\times p \times d}$ is a sample-invariant dictionary (\emph{i.e.}, a set of convolutional filters) representing \textbf{common} local patterns of metal artifacts in different metal-corrupted CT images; $\K_{n}\in\mathbb{R}^{d}$ is a sample-wise weighting coefficient to learn the \textbf{specific} convolutional filter for $\A$ by computing $\mD\ast \K_{n}$ as $\sum^{d}_{i=1} \mD[:,:,i]\odot  \K_{n}[i]$ (see Fig.~\ref{figintro}); $\bm{M}_{n}\in \mathbb R^{H \times W}$ is the coefficients representing the locations for the local patterns of metal artifacts; $p$ is the size of convolutional filters; $d$ is the total number of filters in the dictionary $\mD$; $N$ is the actual number of filters encoding the artifacts $\A$; and the $\otimes$ between $\left(\mD\ast \K \right)$ and $\mM$ is a convolutional operation in tensor form. Specifically, $\left(\mD\ast \K_{n} \right)\in\mathbb{R}^{p\times p}$, $\K \in\mathbb{R}^{d \times N}$, $\left(\mD\ast \K \right)\in\mathbb{R}^{p\times p \times N}$, $\mM\in\mathbb{R}^{H\times W\times N}$, and
% Here, $\mK\in\mathbb{R}^{f\times f\times N}$ and $\mM\in\mathbb{R}^{H\times W\times N}$ are constructed by stacking $\K_{n}$ and $\M_n$ ($n \in\{1, 2, \dots, N\}$), respectively, as:\footnote{Accurately, $\left(\mD\ast \K \right)\in\mathbb{R}^{p\times p\times 1\times N}$. The last dimension of ``1" means single channel input, \emph{i.e.}, grayscale CT images. For simplicity, we omit the last dimension when defining the size of these mathematical symbols including $\Y$, $\X$, $\A$, $\I$, and $\left(\mD\ast \K_{n} \right)$.}
\begin{equation}
\small
\begin{split}
&\mK = \left[\K_{1}, \K_{2}, \cdots, \K_{N}\right],~~\mM = \left[\bm{M}_1, \bm{M}_2, \cdots, \bm{M}_N\right],\\
&\mD \ast \K = \left[\mD\ast \K_{1},\mD\ast \K_{2},\cdots, \mD\ast \K_{N} \right].
\end{split}
\end{equation}
\normalsize

Note that convolutional dictionary model has been verified to be applicable to represent repetitive patterns by existing studies~\cite{wang2021dicdnet}. Compared to these methods, the proposed weighted mechanism in Eq.~\eqref{A} has two main merits for the MAR task: 1) It delivers not only the sample-invariant/common knowledge, but also the sample-wise characteristics for information embedding, which can adaptively encode the prior structure for every CT image; 2) With such a weighted convolutional dictionary, we can choose an $N$ smaller than $d$,\footnote{In all our experiments below, $N=6$ and $d=32$.} which further shrinks the solution space for estimating $\A$ and thereby improves the model generalization (see Sec.~\ref{sec:exp}).

% this weighting-based convolutional dictionary mechanism in Eq.~\eqref{A} is an extended form of convolutional dictionary model~\cite{gu2017joint, zhang2017convolutional,li2018video} which is applicable to represent repetitive local patterns. For metal artifacts,
By substituting Eq.~(\ref{A}) into Eq.~(\ref{e1}), we can derive:
\begin{equation}\label{finaly}
\I\odot\Y=\I\odot\X +\I\odot\left(\left(\mD\ast\mK\right)\otimes \mM\right).
\end{equation}
As shown, our goal is to estimate the sample-wise $\K$, $\mM$, and $\X$ from $\Y$. Note that the non-metal mask $\I$ is pre-known (see Sec.~\ref{sec:exp}) and the common dictionary $\mD$ can be automatically learnt from training data (see Sec.~\ref{sec:net}). With the maximum-a-posterior framework, the optimization problem is:
% sample-invariant
% Actually, we can regard the kernel $\mK$ as a convolutional dictionary~\cite{Huang2015Convolutional} for representing repetitive and similar local patterns of metal artifacts. They deliver the common prior knowledge for describing different artifacts across metal-affected CT images, and thus be automatically learned from the training samples by virtue of the strong learning capability of deep networks (detailed in Sec.~\ref{sec:net}).
% Unlike $\mK$, the feature map $\mM$ is related to the location of artifacts and varies with the metal-corrupted image $\Y$. Hence,  at the testing phase, our goal is to infer the $\mM$ and $\X$ from $\Y$ with the fixed $\mK$. Correspondingly, the optimization problem is formulated as:
% in an end-to-end manner
\begin{equation}\label{o1}
\begin{split}
\min_{{\K,\mM,\X}}&\left\|\I\odot\left(\Y\!-\!\X\!-\!\left(\mD\ast\mK\right)\!\otimes\!\mM\right)\right\|_{F}^{2}\\&~~+\alpha f_{1}(\K)\!+\beta f_{2}(\mM)\!+\!\gamma f_{3}(\X)\\
&\hspace{-10mm} \text{subject to}~~~ \left\|\K_{n}\right\|_{2}=1, n =1,2,\ldots, N,
\end{split}
\end{equation}
where $f_{1}\left(\cdot\right)$, $f_{2}\left(\cdot\right)$,
and $f_{3}\left(\cdot\right)$ are regularization functions, delivering the prior knowledge of $\K$, $\mM$, and $\X$, respectively; $\alpha$, $\beta$, and $\gamma$ are regularization weights.

\subsection{Optimization Algorithm}\label{sec:opt}
Traditional solvers for the problem~(\ref{o1}) often involve complex Fourier and inverse Fourier transforms, which are difficult to integrate into deep networks. In this regard, we design a new optimization algorithm with only simple operators. Concretely, the proximal gradient technique~\cite{beck2009fast} is adopted to alternately update $\K$, $\mM$, and $\X$:
% . The proposed algorithm with only simple computations
% to exactly transform the solution algorithm to network modules

\paragraph{Solving $\K$:} 
From the problem~\eqref{o1}, at the $(t+1)$-th iteration, $\K$ is solved as follows:
\begin{equation}\label{ok}
    \small
    \begin{split}
    &\K^{(t+1)} \!=\! \underset{{\mK}}{\text{argmin}}\left\|\I\!\odot\!\left(\Y\!\!-\!\!\X^{(t)}\!\!-\!\!\left(\mD\ast\mK\right)\otimes\mM^{(t)}\right)\right\|_{F}^{2}\!\!+\!\alpha f_{1}(\K),\\
    & \text{subject to} ~~~\left\|\K_{n}\right\|_{2}=1, n =1,2,\ldots, N,
\end{split}
\end{equation}
\normalsize
and the quadratic approximation~\cite{beck2009fast} of the objective function in the problem (\ref{ok}) is:
\begin{equation}\label{ok2}
    \begin{split}
     \K^{(t+1)}&=\underset{\K \in \Omega}{\text{argmin}}~~~g_{1}\left(\K^{(t)}\!\right) +\frac{1}{2\eta_{1}}\left\|\K-\K^{(t)} \right\|_F^2  \\&\hspace{3mm}+\left\langle  \K-\K^{(t)}, \nabla g_{1}\left(\K^{(t)}\!\right) \right\rangle
    +\alpha f_{1}\left( \K \right),
    \end{split}
\end{equation}
where $\Omega=\{\K|\left\|\K_{n}\right\|_{2}=1, n=1,2,\cdots,N\}$;
$g_{1}\left(\K^{(t)}\right)=\! \left\|\I\odot\left(\Y\!-\!\X^{(t)}\!-\left(\mD\ast\mK^{(t)}\right)\!\otimes\!\mM^{(t)}\right)\right\|_{F}^{2}$; $\eta_{1}$ is the stepsize. Hence, we can get the following equation:
\begin{equation}
    \small
    \begin{split}
    &\frac{\partial g_{1}\left(\K^{(t)}\right)}{\partial \K_{n}}\!=\!\left(\!U^{f}_{3}\!\left(\!\mathcal{D}\otimes^{d} \bm{M}_{n}^{(t)}\right)\!\right)\!\\&\quad\quad\quad\text{vec}\!\left(\I\odot\left(\left(\mD\ast\mK^{(t)}\right) \otimes \mM^{(t)}\!+\!\X^{(t)}\!-\!\Y\right)\right),
    \end{split}
\end{equation}
\normalsize
where $\otimes^{d}$ is the depth-wise convolutional computation; $U^{f}_{3}(\cdot)$ denotes the unfolding operation at the $3^{rd}$ mode; and vec($\cdot$) denotes the vectorization.

%With $\mathcal{D}\otimes^{d} \bm{M}_{n}^{(t)}$, the result is with the size as $H\times W \times d$. Then, with , \emph{i.e.}, $U^{f}_{3}(\cdot)$, the result  has the size of $d \times HW$.
%$(\cdot)^{T}$ represents the matrix transpose operation.

Clearly, Eq.~\eqref{ok2} can be equivalently written as:
\begin{equation}\label{ok3}
\small
  \K^{(t+1)}=\underset{\K \in \Omega}{\text{argmin}} \frac{1}{2} \left\| \K \!-\!\! \left(\! \K^{(t)}\!\!-\! \eta_{1}\nabla g_{1}\left(\K^{(t)}\right)  \!\right) \right\|_F^2 \!+ \alpha\eta_{1} f_{1}\left( \K \right).
\end{equation}
For general prior $f_{1}(\cdot)$~\cite{donoho1995noising}, Eq.~\eqref{ok3} is derived as:
\begin{equation}\label{ok4}
\K^{(t+1)}=
  \mbox{prox}_{\alpha\eta_{1}}\!\!\left(\K^{(t+0.5)}\right),
\end{equation}
where $\K^{(t+0.5)}=\K^{(t)}
  - \eta_{1}\nabla g_{1}\left(\K^{(t)} \right)$;
$\nabla g_{1}\left(\K^{(t)}\right) \!\!= \!\!\left[\!
\frac{\partial  g_{1}\left(\K^{(t)}\right)}{\partial \K_{1}},
\frac{\partial  g_{1}\left(\K^{(t)}\right)}{\partial \K_{2}},\ldots, \frac{\partial  g_{1}\left(\K^{(t)}\right)}{\partial \K_{N}}\!\right]$; $\mbox{prox}_{\alpha\eta_{1}}\left(\cdot\right)$ is a proximal operator which is related to $f_{1}\left(\cdot\right)$. The constraint space $\Omega$ can be achieved by introducing a normalization operation into $\mbox{prox}_{\alpha\eta_{1}}\left(\cdot\right)$ (see Sec.~\ref{sec:net}).

\paragraph{Solving $\mM$:} Similarly, at the $(t+1)$-th iteration, 
$\mM$ can be updated by solving the quadratic approximation of the subproblem with respect to $\mM$ as:
\begin{equation}\label{om2}
\small
\begin{split}
\mM^{(t+1)}=\underset{\mM}{\text{argmin}} &\frac{1}{2} \left\| \mM -\left(\! \mM^{(t)}- \eta_{2}\nabla g_{2}\left(\mM^{(t)}\right)  \right) \right\|_F^2\\&\vspace{6mm} + \beta\eta_{2} f_{2}\left( \mM \right),
\end{split}
\end{equation}
\normalsize
where $\eta_{2}$ is the stepsize and $g_{2}\left(\mM^{(t)}\right)\!=\! \left\|\I\odot\left(\Y\!-\!\X^{(t)}\!-\!\left(\mD\ast\mK^{(t+1)}\right)\!\otimes\!\mM^{(t)}\right)\right\|_{F}^{2}$. Similarly, the solution of problem~(\ref{om2}) is deduced as:
\begin{equation}\label{om3}
  \mM^{(t+1)}=\mbox{prox}_{\beta\eta_{2}}\left(\mM^{(t+0.5)}\right),
\end{equation}
where $\mbox{prox}_{\beta\eta_2}(\cdot)$ is a proximal operator related to the regularization function $f_{2}(\cdot)$ about $\mM$; $\mM^{(t+0.5)}=\mM^{(t)} \!-\! \eta_{2}\nabla g_{2}\left(\mM^{(t)}\right)$; $\nabla g_{2} \left(\mM^{(t)}\right)=\left(\mD\ast\mK^{(t+1)}\right)\otimes^{T}\!\left(\I\odot\left(\left(\mD\ast\mK^{(t+1)}\right) \otimes \mM^{(t)}\!+\!\X^{(t)}\!-\!\Y\right)\right)$; and
$\otimes^T$ is the transposed convolution operation.

\begin{figure*}[t]
  \begin{center}
     \includegraphics[width=0.90\linewidth]{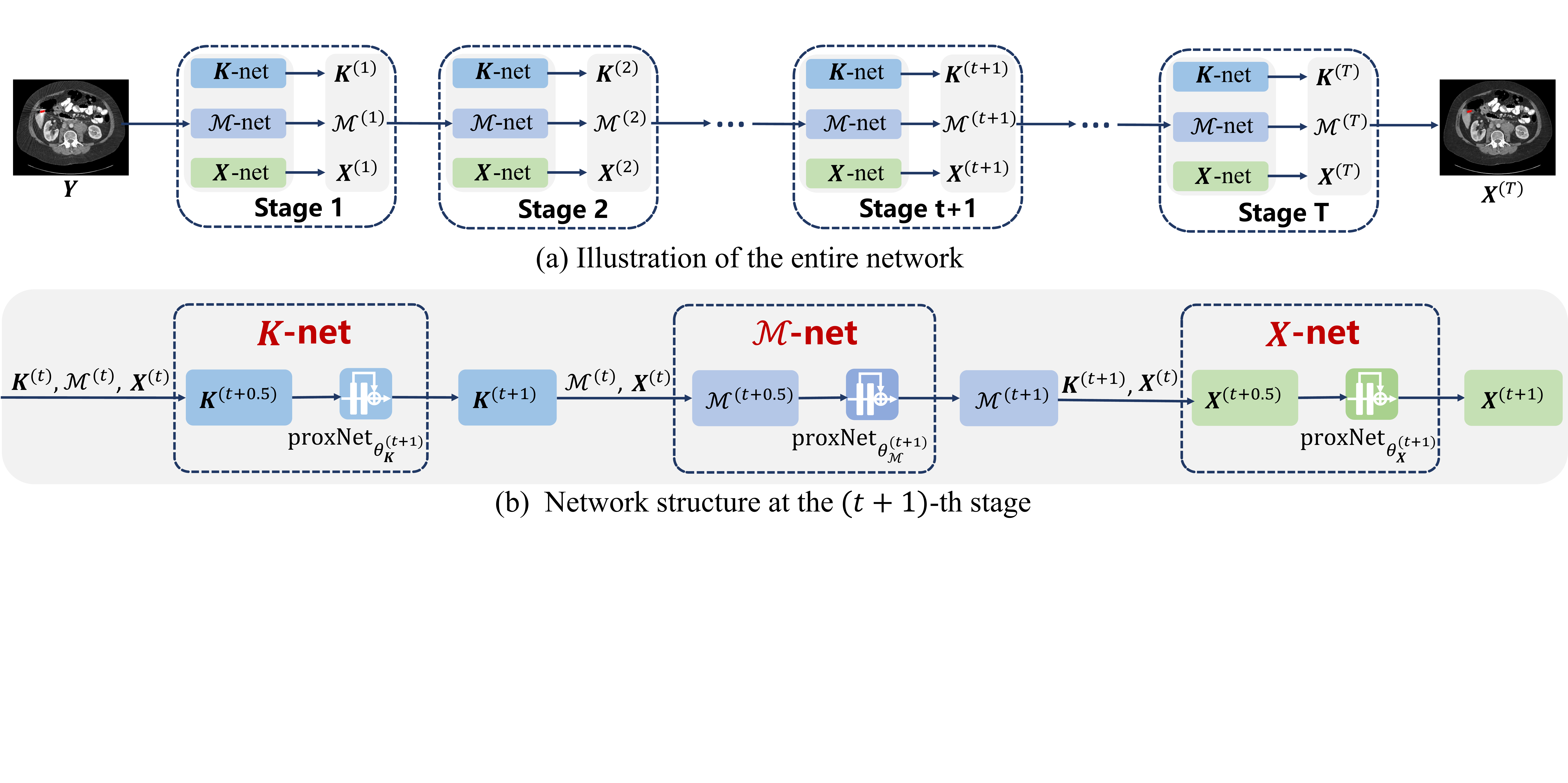}
  \end{center}
  \vspace{-4mm}
     \caption{(a) The proposed ACDNet consists of $T$ stages. (b) The detailed structure at any stage where $\K^{(t+1)}$, $\mM^{(t+1)}$, and $\X^{(t+1)}$ are successively updated by $\K$-net, $\mM$-net, and $\X$-net, respectively, based on Eqs.~(\ref{ok4}), (\ref{om3}), and (\ref{ox3}).}
  \label{fignet}
    \vspace{-2mm}
\end{figure*}

\paragraph{Solving $\X$:} 
Given $\K^{(t+1)}$ and $\mM^{(t+1)}$, the quadratic approximation of the subproblem about $\X$ is derived as:
\small
\begin{equation}\label{ox2}
\X^{(t+1)}\!=\!\underset{\X}{\text{argmin}}\frac{1}{2}\!\left\| \X \!-\! \left(\! \X^{(t)} \!-\! \eta_{3}\nabla g_{3}\left(\X^{(t)}\right)  \!\right) \right\|_F^2 \!\!+\! \gamma\eta_{3} f_{3}\! \left( \X \right),
\end{equation}
\normalsize
where $g_{3}\left(\!\X^{(t)}\!\right)\!=\! \left\|\I\!\odot\!\left(\!\Y\!\!-\!\!\X^{(t)}\!\!-\!\!\left(\mD\!\ast\!\mK^{(t+1)}\right)\!\otimes\!\mM^{(t+1)}\right)\!\right\|_{F}^{2}$.
Then, the updating rule of $\X$ is written as:
% \small
% \begin{equation}\label{updateb}
% \begin{split}
%   &\X^{(s)} \!=\!\\
%   &\mbox{prox}_{\beta\eta_{2}}\!\!\left(\!\left(\bm{1}-\eta_2\I\right) \odot\X^{(s-1)}
%   \!\!+\!\eta_{2}\I\odot\!\left(\!\Y\! - \mK \!\otimes\! \mM^{(s)}\!\right)\!\right)\!,
% \end{split}
% \end{equation}
% \normalsize
\begin{equation}\label{ox3}
\X^{(t+1)} =\mbox{prox}_{\gamma\eta_{3}}\left(\X^{(t+0.5)}\right),
\end{equation}
\normalsize
% $X^{(t+0.5)}=\left(\bm{1}-\eta_3\I\right) \odot\X^{(t)}
%   \!\!+\!\eta_{3}\I\odot\!\left(\!\Y\! - \left(\mD\ast\mK^{(t+1)}\right) \!\otimes\! \mM^{(t+1)}\!\right)$
where $\X^{(t+0.5)}\!=\!\left(\bm{1}\!-\!\eta_3\I\right) \odot\X^{(t)}
  +\eta_{3}\I\odot\left(\!\Y\! \!-\!\! \left(\mD\ast\mK^{(t+1)}\right) \!\otimes\! \mM^{(t+1)}\!\right)$; $\mbox{prox}_{\gamma\eta_3}(\cdot)$ is dependent on the prior function $f_{3}(\cdot)$ about $\X$.
 %and $\bm{1}$ is a constant matrix with all elements as 1.

Eqs.~\eqref{ok4},~\eqref{om3}, and~\eqref{ox3} constitute the entire iterative process for solving the problem~\eqref{o1}. As seen, the proposed algorithm only contains simple operators,  which makes it easier to accordingly build the deep network framework upon the algorithm. Note that $\mbox{prox}_{\alpha\eta_1}(\cdot)$, $\mbox{prox}_{\beta\eta_2}(\cdot)$, and $\mbox{prox}_{\gamma\eta_3}(\cdot)$ are implicit operators, which are automatically learnt from training data by virtue of the powerful prior fitting capability of deep networks. This manner has been fully used and widely validated to be effective and helpful to explore interpretable knowledge by recent studies (\emph{e.g.}. \cite{wang2021rcdnet,xie2020mhf}). The details are in Sec.~\ref{sec:net}.

%$\nabla h\left(\!\B^{(s-1)}\!\right) = \I\odot\left(\mK\otimes \mM^{(s)}+ \I\odot\B^{(s-1)}- \I\odot\Y\right)$

\section{Network Design and Analysis}\label{sec:net}
Due to the specific design of joint model-driven (\emph{i.e.}, prior knowledge) and data-driven (\emph{i.e.}, deep networks) frameworks, deep unfolding techniques have achieved great success in computer vision tasks, such as
single image rain removal~\cite{wang2021rcdnet} and low-light image enhancement~\cite{liu2022low}. Inspired by this, we specifically design an adaptive convolutional dictionary network (ACDNet) for the MAR task by unfolding the iterative algorithm in Sec.~\ref{sec:opt} into the corresponding network structure.

Fig.~\ref{fignet} (a) displays the entire network structure with $T$ stages, which correspond to $T$ iterations of the optimization algorithm in Sec.~\ref{sec:opt}. For every stage, our network consists of three sub-nets, \emph{i.e.}, $\K$-net, $\mM$-net, and $\X$-net corresponding to addressing the three subproblems, \emph{i.e.,}, solving $\K$, $\mM$, and $\X$, respectively.
Fig.~\ref{fignet} (b) shows the detailed network connections at every stage, which are constructed by sequentially unfolding the iterative rules, \emph{i.e.}, Eq.~\eqref{ok4}, Eq.~\eqref{om3}, and Eq.~\eqref{ox3}, respectively. With the unfolding operations, every network module corresponds to the specific iterative step of the proposed optimization algorithm. Thus, the entire network framework has a clear physical interpretability. Next we present the information of sub-nets in detail:
% Specifically, at every stage, the network details are described as:

\paragraph{$\K$-net:}
At the $(t+1)$-th stage, $\K^{(t+0.5)}$ is computed and fed to {$\text{proxNet}_{\theta_{\K}^{(t+1)}}(\cdot)$} to execute the operator $\mbox{prox}_{\alpha\eta_{1}}(\cdot)$. Then, the updated weighting coefficient is:
{{$\K^{(t+1)} = \text{proxNet}_{\theta_{\K}^{(t+1)}}\left(\K^{(t+0.5)}\right)$}}, where {$\text{proxNet}_{\theta_{\K}^{(t+1)}}(\cdot)$} is a residual structure, \emph{i.e.}, [\emph{Linear+ReLU+Linear+Skip Connection+Normalization at the dimension $d$}].

\paragraph{$\mM$-net:}
Given $\K^{(t+1)}$, $\mM^{(t)}$, and $\X^{(t)}$,
we can compute {{$\mM^{(t+1)} = \text{proxNet}_{\theta_{\mM}^{(t+1)}}\left(\mM^{(t+0.5)}\right)$}}, where {$\text{proxNet}_{\theta_{\mM}^{(t+1)}}(\cdot)$} is the unfolding form of $\mbox{prox}_{\beta\eta_{2}}(\cdot)$---ResNet with three [{\normalsize{\emph{Conv+BN+ReLU+Conv+BN+Skip Connection}}}] residual blocks.

\paragraph{$\X$-net:} 
Similarly, given $\K^{(t+1)}$, $\mM^{(t+1)}$, and $\X^{(t)}$,
the artifact-reduced CT image can be updated by
$\X^{(t+1)} =\mbox{proxNet}_{\theta_{\X}^{(t+1)}}\left(\X^{(t+0.5)}\right)$, where $\text{proxNet}_{\theta_{\X}^{(t+1)}}(\cdot)$ has the same network structure to $\text{proxNet}_{\theta_{\mM}^{(t+1)}}(\cdot)$.

With $T$-stage optimization, we can get the final CT image $\X^{(T)}$. All the involved parameters are $\{\theta_{\K}^{(t)},\theta_\mM^{(t)},\theta_{\X}^{(t)}\}_{t=1}^{T}$, $\{\eta_{i}\}_{i=1}^{3}$, and the common dictionary $\mD$. They can be automatically learnt from training samples in an end-to-end manner. Note that $\mD$ is achieved by a common convolutional layer. More details, including the initialization ($\K^{(0)}$, $\mM^{(0)}$, $\X^{(0)}$), are given in \textit{Supplemental Material (SM)}.

\begin{figure*}[t]
  \begin{center}
  %\vspace{-3mm}
     \includegraphics[width=0.91\linewidth]{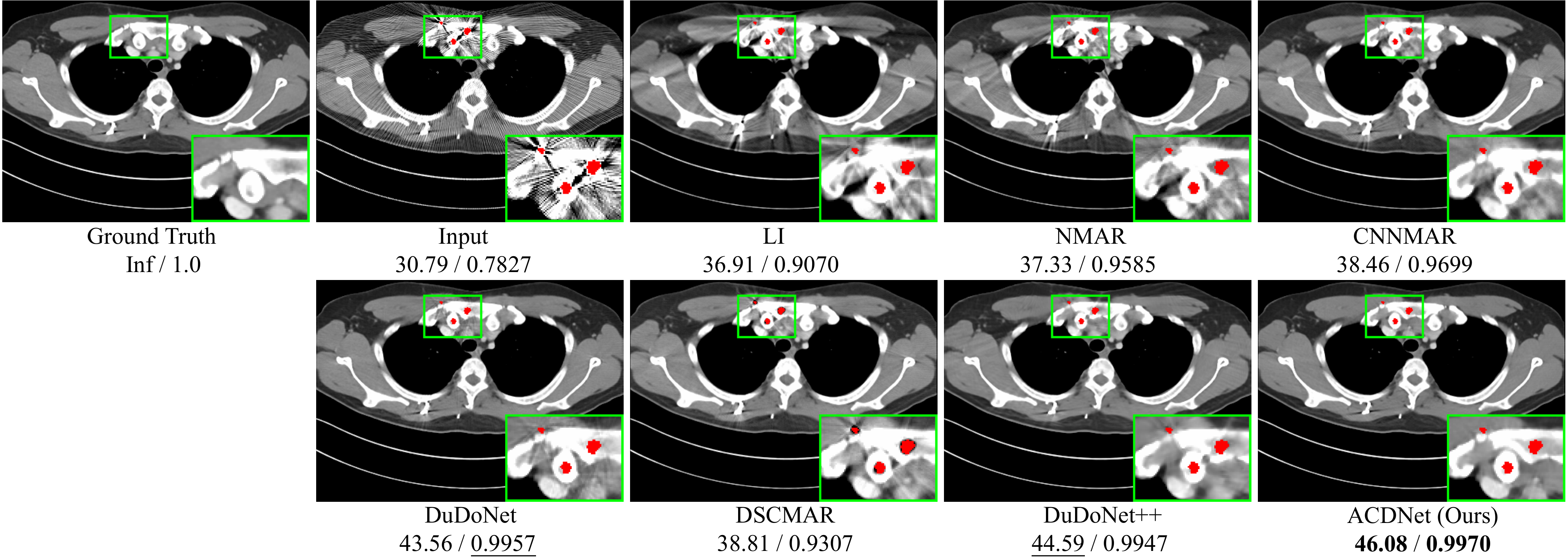}
  \end{center}
  \vspace{-4mm}
     \caption{Performance comparison of different MAR approaches on a metal-corrupted CT image selected from the synthesized DeepLesion data. PSNR (dB)/SSIM below is for reference. The red pixels stand for metallic implants.} % (Training-testing domain match)
  \label{figsyn}
  \vspace{-1mm}
\end{figure*}
 \begin{table*}[t]
\centering
\small
%Bold and bold italic denote top $1^{\text{st}}$ and $2^{\text{nd}}$ ranks, respectively.
% $^{\text{*}}$ means we adopt the pre-trained model released by the authors \Cite{wei2019semi}. (Training-testing domain match)
%\Xegin{tabular}{@{}c|c@{}c|c|c|c|c|c|c@{}}
\setlength{\tabcolsep}{4.1pt}
    \begin{tabular}{l|c|c|c|c|c|c}
    %\hline
    \hline
    Methods    & \multicolumn{5}{c|}{ Large Metal \quad \quad   \quad\quad  $\longrightarrow$    \quad   \quad  \quad  \quad \quad  Small Metal}                & Average      \\
    \hline
    Input             &24.12/0.6761              &26.13/0.7471              &27.75/0.7659               &28.53/0.7964              &28.78/0.8076              &27.06/0.7586              \\
    %BHC~\cite{verburg2012ct} &26.31/0.8106 &27.52/0.8583 &29.17/0.8973 &29.61/0.9208 %&29.71/0.9251&28.46/0.8824\\
    LI~\cite{kalender1987reduction}              &27.21/0.8920              &28.31/0.9185              &29.86/0.9464              &30.40/0.9555              &30.57/0.9608              &29.27/0.9347\\
    NMAR~\cite{meyer2010normalized}              &27.66/0.9114              &28.81/0.9373              &29.69/0.9465              &30.44/0.9591              &30.79/0.9669              &29.48/0.9442              \\
  CNNMAR~\cite{zhang2018convolutional}            &28.92/0.9433  &29.89/0.9588  & 30.84/0.9706             &31.11/0.9743              &31.14/0.9752              &30.38/0.9644              \\
DuDoNet~\cite{lin2019dudonet}           & 29.87/0.9723 & 30.60/0.9786 & 31.46/0.9839  & 31.85/0.9858 & 31.91/0.9862 & 31.14/0.9814 \\
DSCMAR~\cite{yu2020deep}           & 34.04/0.9343 & 33.10/0.9362 & 33.37/0.9384  & 32.75/0.9393 & 32.77/0.9395 & 33.21/0.9375 \\
    DuDoNet++~\cite{lyu2020dudonet++}         &\underline{36.17}/\underline{{0.9784}}& \underline{{38.34}}/{\underline{0.9891}} &\underline{{40.32}}/\underline{{0.9913}}
    &\underline{41.56}/\underline{{0.9919}}& \underline{{42.08}}/\underline{{0.9921}}
    &\underline{39.69}/\underline{{0.9886}}\\
    ACDNet (Ours)  &\textbf{37.91}/\textbf{0.9872} &\textbf{39.30}/\textbf{0.9920} & \textbf{41.14}/\textbf{0.9949}  & \textbf{42.43}/\textbf{0.9961} & \textbf{42.64}/\textbf{0.9965} &\textbf{40.68}/\textbf{0.9933}\\
    \hline
    \end{tabular}
    \vspace{-1mm}
    \caption{Average PSNR (dB)/SSIM of different MAR methods on the synthesized DeepLesion data.}
\label{tabsyn}
\vspace{-4mm}
\end{table*}

\paragraph{\emph{ProxNet:}} Following recent great deep unfolding-based works~\cite{wang2021indudonet,xie2020mhf}, we also set proximal operator with ResNet. Although it is very hard to inversely derive the form of regularizer by integrating ResNet function due to its complicated form, but as has comprehensively substantiated by previous research, it does be helpful to explore insightful structural prior. Besides, adopting such learning-based manner to automatically learn implicit regularizers is more flexible, avoiding hand-crafted prior design. 

\paragraph{\emph{Interpretability:}} Unlike most of current deep MAR networks, which are heuristically built based on the off-the-shelf network blocks, ACDNet is naturally constructed under the guidance of the optimization algorithm with careful data fidelity term design. In this regard, the entire network integrates the interpretability of model-based methods. {\color{black}{Besides, such interpretability is visually validated by Fig. 3 of \textit{SM}.}}

\paragraph{\emph{Remark:}} Deep unfolding technique
is a general tool for network design. To apply it, the key challenge is how to elaborately design model and optimization algorithm to make it work for specific applications. For the MAR task, we have specifically made some substantial ameliorations: 1) The adaptive prior of artifacts is incorporated into data fidelity term which can finely guide the network learning and boost the generalization ability; 2) Unlike other solvers with complicated computations (\emph{e.g.}, Fourier transformation and matrix inversion), the proposed optimization algorithm is elaborately designed which only contains simple operators and makes the unfolding process easy and proper; 3) Compared to current popular dual-domain MAR methods, ACDNet only uses CT image domain data, which is more friendly to practical applications where the projection data is difficult to acquire. Besides, ACDNet can be easily integrated into current computer-aided diagnosis systems as a plug-in module.  

\paragraph{Loss Function.}
With supervision on the extracted artifact $\A^{(t)}$ and CT image $\X^{(t)}$ at every stage, the training loss is:
\begin{equation}\label{Loss}
    \small
    \begin{split}
    %\vspace{-1mm}
     \mathcal{L}& \!= \!\!\sum_{t=0}^{T}\mu_{t}\I\!\odot\!\left\|\X\!\!-\!\!\X^{(t)} \right\|_F^2\! \!+\!\omega_{1}\!\left(\sum_{t=0}^{T}\mu_{t}\I\!\odot\!\left\|\X\!\!-\!\!\X^{(t)} \right\|_1\!\right)\\& +\omega_{2}\left(\sum_{t=0}^{T}\mu_{t}\I\odot\left\|\Y-\X-\A^{(t)} \right\|_1\right),
    %\vspace{-1mm}
    \end{split}
\end{equation}
\normalsize
where $\X$ is the clean (\emph{i.e.}, ground truth) CT image. In all experiments, $\mu_{T}$ is set to 1; $\mu_{t}\,\,\,(t=0,1,\cdots,T-1)$ is 0.1;
% to supervise the outputs at every middle stage.
$\omega_{1}$ and $\omega_{2}$ are empirically set to $5\times 10^{-4}$.

\paragraph{Implementation Details.}
ACDNet is optimized through Adam optimizer based on PyTorch. The framework is trained on an NVIDIA Tesla V100-SMX2 GPU with a batch size of 32. The initial learning rate is $2\times10^{-4}$ and divided by 2 at epochs [50, 100, 150, 200]. The total number of epochs is 300. The size of input image patch is $64\times64$ pixels and it is randomly flipped horizontally and vertically. More explanations are included in \textit{SM}.

% \begin{figure*}[!t]
%   \begin{center}
%   %\vspace{-3mm}
%      \includegraphics[width=0.9\linewidth]{PIC/ijcai_modelverif.pdf}
%   \end{center}
%   \vspace{-4mm}
%      \caption{Visualization of artifact-reduced image $\X^{(t)}$ and artifact $\A^{(t)}$ extracted by our ACDNet at the stage $t$. The total stage $T$ is 10 and $\A$ is the groudtruth artifact, expressed as $\left(\Y-\X\right)\odot\I$.} % (Training-testing domain match)
%   \label{figseg}
%   \vspace{-0mm}
% \end{figure*}

\begin{figure*}[t]
  \begin{center}
  %\vspace{-3mm}
     \includegraphics[width=0.91\linewidth]{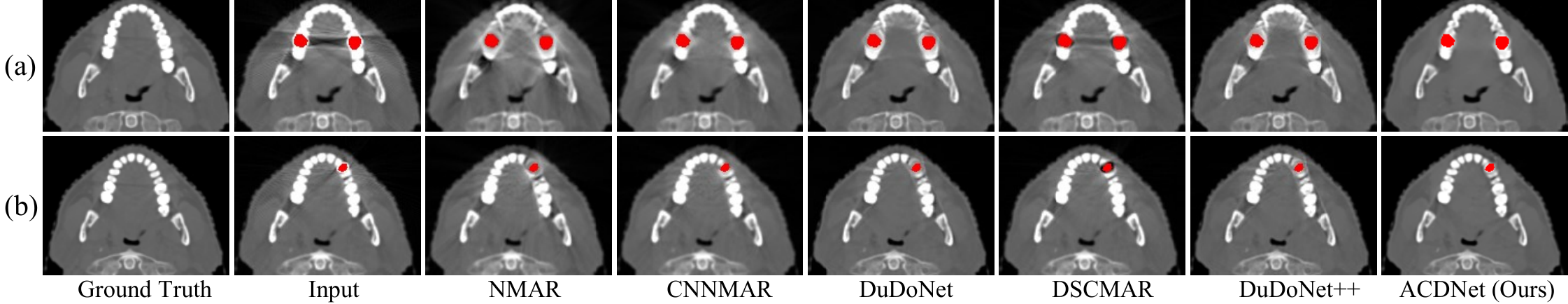}
  \end{center}
  \vspace{-4mm}
     \caption{Generalization results.  Artifact-reduced results on the synthesized dental CT images with different metallic implants.} % (Training-testing domain match)
  \label{figdental}
    \vspace{-3mm}
\end{figure*}
\begin{figure*}[!t]
  \begin{center}
  %\vspace{-3mm}
     \includegraphics[width=0.91\linewidth]{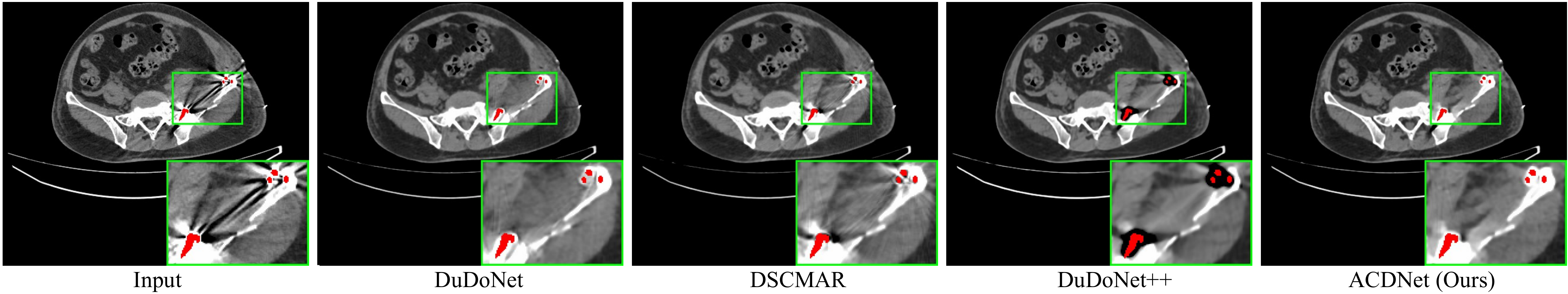}
  \end{center}
  \vspace{-4mm}
     \caption{Generalization results. Performance comparison on a real clinical metal-affected CT image from CLINIC-metal.}
  \label{figclinic}
  \vspace{-3mm}
\end{figure*}
\begin{figure*}[!t]
  \begin{center}
  %\vspace{-3mm}
     \includegraphics[width=0.905\linewidth]{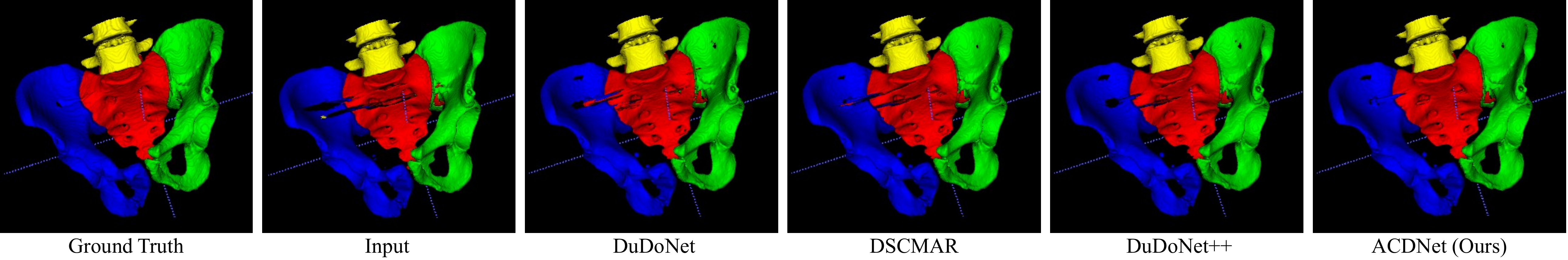}
  \end{center}
  \vspace{-4mm}
     \caption{Downstream segmentation results of an artifact-reduced CLINIC-metal volume reconstructed by different MAR approaches.} % (Training-testing domain match)
  \label{figseg}
  \vspace{-3mm}
\end{figure*}

\vspace{-0mm}
\section{Experiments}\label{sec:exp}
In this section, we conduct extensive experiments to validate the effectiveness of our method.\footnote{More results, including interpretability verification, ablation study, and P-values analysis are given in \textit{SM}.}
%and compare its performance with state-of-the-art (SOTA) MAR approaches.%\footnote{More experimental results are given in \textit{Supplemental Material}.}
\subsection{Dataset \& Experimental Setting}
\vspace{0mm}
\noindent\textbf{Synthesized Data.}
Following the simulation procedure in~\cite{yu2020deep}, we randomly choose 1,200 clean CT images from the public DeepLesion dataset~\cite{yan2018deep} and collect 100 metal masks from~\cite{zhang2018convolutional} to synthesize the paired clean/metal-corrupted CT images. Specifically, 90 metals together with 1000 clean CT images for training and 10 ones together with the remaining 200 clean CT images for testing. Similar to~\cite{lin2019dudonet,lyu2020dudonet++}, we sequentially take every two testing metal masks as one group for performance evaluation. In addition, to evaluate the cross-body-site generalization performance, clean dental CT images~\cite{yu2020deep} are adopted and the corresponding metal-corrupted images are generated under the same simulation protocol on DeepLesion.
% and adopt the aforementioned simulation protocol 
% to generate paired CT images for testing evaluation. 

\vspace{1mm}
\noindent\textbf{Clinical Data.} 
A public clinical dataset, \emph{i.e.}, CLINIC-metal~\cite{liu2020deep}, which contains 14 metal-corrupted volumes with pixel-wise annotations of multiple bone structures (\emph{i.e.,} sacrum, left hip, right hip, and lumbar spine), is used for evaluation. Following~\cite{yu2020deep}, the clinical metal masks are segmented with thresholding of 2500 HU.

\vspace{1mm}
\noindent\textbf{Baselines.} Representative MAR methods are used, including traditional LI~\cite{kalender1987reduction} and NMAR~\cite{meyer2010normalized}, learning-based CNNMAR~\cite{zhang2018convolutional}, DuDoNet~\cite{lin2019dudonet}, DSCMAR~\cite{yu2020deep}, and DuDoNet++~\cite{lyu2020dudonet++}.

\vspace{1mm}
\noindent\textbf{Evaluation Metric.}
We adopt the PSNR/SSIM for quantitative comparison on synthesized data and only visual comparison on clinical data due to the lack of clean CT images.
%\subsection{Model Verification}

\subsection{Performance Evaluation}
\paragraph{DeepLesion Data.} Fig.~\ref{figsyn} presents the visual comparison on an image from synthesized metal-corrupted DeepLesion. As shown, our approach significantly removes the artifacts and finely recovers evident details. The average PSNR/SSIM scores on the entire DeepLesion dataset are listed in Table~\ref{tabsyn}. It can be observed that ACDNet consistently achieves the best scores with varying sizes of metallic implants.
% , thus achieving better visual quality and higher PSNR/SSIM scores. Table~\ref{tabsyn} lists the average PSNR/SSIM on the entire DeepLesion data. With the metal size varying from large to small, the proposed method consistently outperforms other comparison methods. This confirms the effectiveness and generality of our method.
\paragraph{Dental Data.}
Fig.~\ref{figdental} shows the MAR visual results on the synthesized dental CT images with different metallic implants, where deep MAR methods are trained on the synthesized DeepLesion (focusing on abdomen and thorax) and directly tested on the dental CT images. Due to the regularization with the explicit WCD model, ACDNet can more accurately identify the artifacts and accomplish the better reconstruction of artifact-reduced CT images. The results show the excellent cross-body-site generalizability of our method.
%\footnote{More results are included in \textit{SM}.\label{foot1}}

% From the comparisons, we can easily observe that with the explicit regularization~\eqref{A}, our method can more accurately identify the metal artifacts and accomplish the better reconstruction of artifact-reduced CT images. This substantiates the good cross-site (from abdomen/thorax CT to dental CT) generalization capability of our method.
%\paragraph{Clinical Data.}

%\vspace{1mm}
\paragraph{Clinical Data.} Fig.~\ref{figclinic} presents the MAR comparison of generalization results on a clinical metal-corrupted CT image. Our ACDNet finely preserves more bone structures and removes more artifacts.  Fig.~\ref{figseg} displays the downstream segmentation results of an artifact-reduced CLINIC-metal volume, which are reconstructed by different MAR approaches. The proposed ACDNet is significantly superior to other approaches, which reveals its excellent potential for clinical applications. Due to the limited space, we only provide the results of the latest MAR methods.%\footref{foot1}

% Table~\ref{tabclinic} reports the segmentation accuracy of artifact-reduced CLINIC-metal CT images, which are recovered by different MAR approaches, on the downstream pelvic multi-bone task. The proposed framework is significantly superior to other listed approaches, which reveals its excellent potential for clinical applications. Due to the limited space, we only provide the results of the latest MAR methods. 

\begin{table}[!t]
\centering
%\vspace{1mm}
\small
\setlength{\tabcolsep}{5.5pt}
    \begin{tabular}{l|c|c}
    \hline
    Methods    &\#Parameters   & Testing Time\\
    \hline
    DuDoNet~\cite{lin2019dudonet} &25,834,251 &0.4225 \\
    DSCMAR~\cite{yu2020deep}    &25,834,251      &0.3638\\
    DuDoNet++~\cite{lyu2020dudonet++} & 25,983,627 & 0.8062 \\
    ACDNet (Ours) &\textbf{1,602,809} & \textbf{0.3138} \\
    \hline
    \end{tabular}
        \vspace{-1mm}
    \caption{The number of network parameters and average testing time (seconds) computed on 2000 images with size $416\times416$ pixels on an NVIDIA Tesla V100-SMX2 GPU.}
    \label{tabtime}
    \vspace{-3mm}
\end{table}

\paragraph{Computation Efficiency.}
Table~\ref{tabtime} reports that our method has fewer network parameters (\emph{i.e.,} storage cost) and computation consumption, compared to other methods.
% Table~\ref{tabtime} has shown that compared to the latest DuDoNet++, our ACDNet has lower storage cost (1.6M vs 26M) and higher inference speed (0.31s vs 0.81s). The computation cost is comparable (148.45GFLOPs vs 133.91GFLOPs). 

\vspace{-1mm}
\section{Conclusion and Future Work}
% In this paper, we explored the physical characteristics of metal-corrupted CT images and encoded the prior knowledge (\emph{e.g.}, local repetitive streaking patterns) of metal artifacts as an explicit weighting-based convolutional dictionary model, which simultaneously considered the sample-invariant common patterns and sample-wise specific structures. Based on the proximal gradient technique, we derived a simple solution algorithm to easily construct the corresponding unfolding network. Every network module in such an unfolding network has the clear physical meanings for the specific MAR task. Naturally, the entire network framework is explicitly integrated with prior information of metal artifacts. Experiments executed on synthesized and clinical data have comprehensively verified the excellent generalization ability and high computational efficiency of our method. 
In this paper, for the MAR task, we have proposed a weighted convolutional dictionary model to explicitly and adaptively encode the structural prior of metal artifacts for every input image. Then, by unfolding a simple-yet-effective optimization algorithm, we have easily built the deep network, which integrated the advantages of both model-based methods and learning-based methods. Experiments executed on three public datasets verified the excellent generalizability and excellent computational efficiency of our method. 

Following current state-of-the-art (SOTA) MAR methods, we adopt the thresholding manner to simply segment the metal for clinical data. An unsatisfactory threshold possibly makes tissues be wrongly regarded as metals and most MAR methods would fail to recover image details. Although our method consistently achieve good performances for different sizes of metals and show the robustness, as a direction to further boost the performance of our framework, the joint optimization of automated metal localization and MAR is worthwhile to further explore. Besides, like SOTA MAR methods, we process CT images slice by slice for fairness. It would be very meaningful to do 3D prior modeling for the MAR task.

% CT images are usually reconstructed slice by slice with
% filtered back-projection, while direct 3D reconstruction is
% available only on advanced scanners. Therefore, most current
% SOTA MAR methods as well as ours exploit 2D networks to
% process CT images slice by slice. The 3D prior modeling is
% very meaningful to this field.

% The thresholding method is widely adopted in the current SOTA methods.
% Nevertheless, our method is observed to be robust to metal segmentation. Our method uses only CT image and there is no necessary connection between metal segmentation and artifact modelling; 2) Our method consistently achieves better performance under different types of metals. Smaller thresholding possibly causes that more tissues are mistakenly regarded as metal regions and then the reconstructed images would lose mass content. Analysis of the thresholding method is beyond our focus and we will do specific exploration in the future.

%\clearpage

% \noindent\textbf{Ethical Statement} There are no ethical issues.

% \vspace{-4mm}
\section*{Acknowledgments}
This work was founded by the China NSFC project under contract U21A6005, the Macao Science and Technology Development Fund under Grant 061/2020/A2, the major key project of PCL under contract PCL2021A12, the Key-Area Research and Development Program of Guangdong Province, China (No. 2018B010111001), National Key R\&D Program of China (2018YFC2000702), the Scientific and Technical Innovation 2030-``New Generation Artificial Intelligence'' Project (No. 2020AAA0104100).

% \vspace{-4mm}
% \section*{Ethical Statement} 
% There are no ethical issues.

%% The file named.bst is a bibliography style file for BibTeX 0.99c
\vspace{-2mm}
\bibliographystyle{named}
\bibliography{ijcai22}

\appendix    
\setcounter{table}{0} 
\setcounter{figure}{0}
\setcounter{equation}{0}

 \begin{figure*}[t]
  \begin{center}
     \includegraphics[width=1\linewidth]{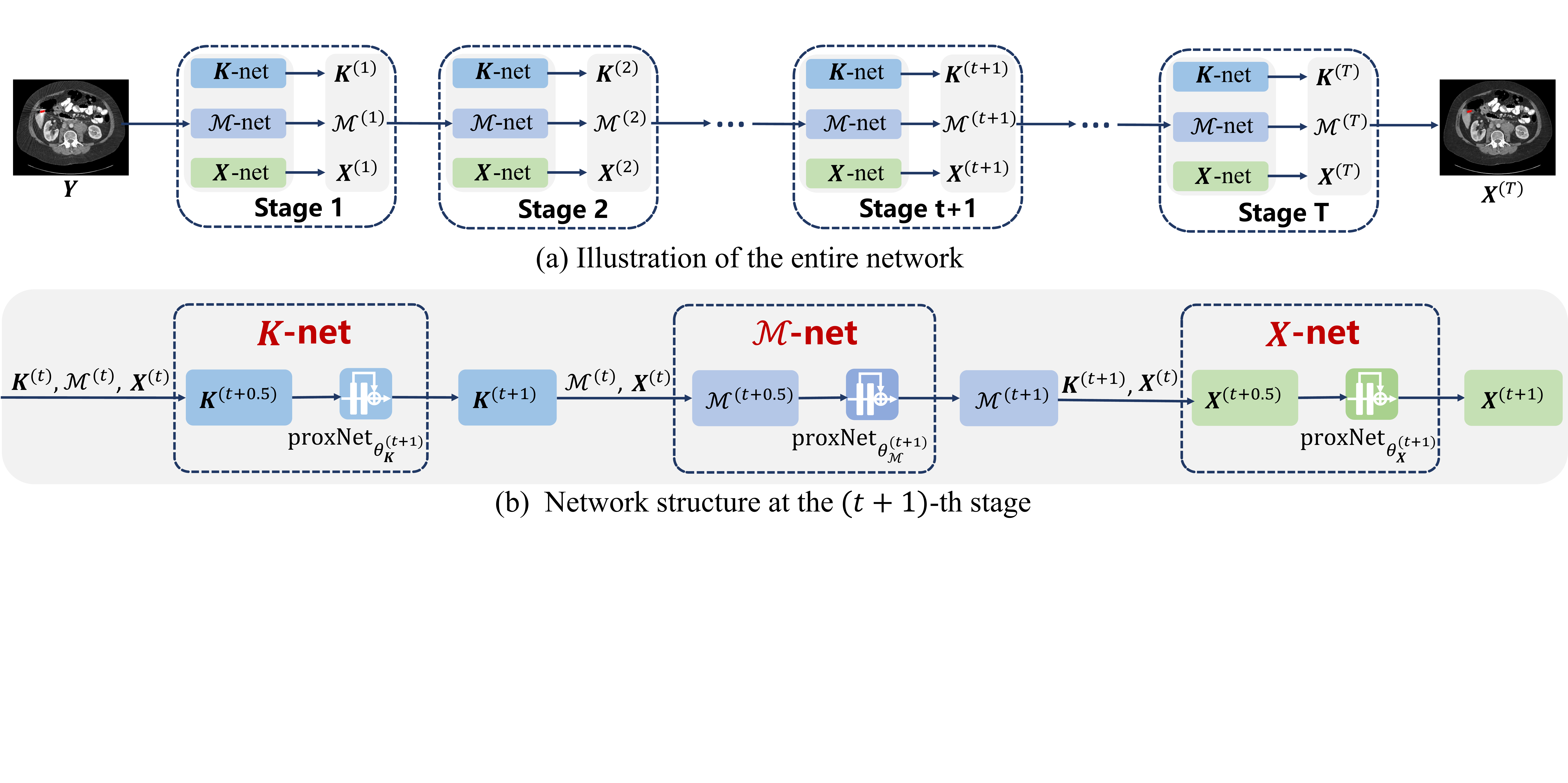}
  \end{center}
  \vspace{-1mm}
     \caption{(a) The proposed network architecture consists of $T$ stages. At every stage, it is sequentially composed of $\K$-net, $\mM$-net, and $\X$-net. (b) The detailed structure at any stage where $\K^{(t+1)}$, $\mM^{(t+1)}$, and $\X^{(t+1)}$ are successively updated by $\K$-net, $\mM$-net, and $\X$-net, respectively, based on the iterative algorithm as Eqs.~(10), (12), and (14) of the main text.}
  \label{fignet}
  \vspace{-2mm}
\end{figure*}
\section*{Supplementary Material}

% Single author syntax
% \author{
%     Author Name
%     \affiliations
%     Affiliation
%     \emails
%     pcchair@ijcai-22.org
% }

% \author{
% Paper ID 5412
% }

\section{More Details about Network Structure}

\begin{figure}[t]
  \begin{center}
     \includegraphics[width=1\linewidth]{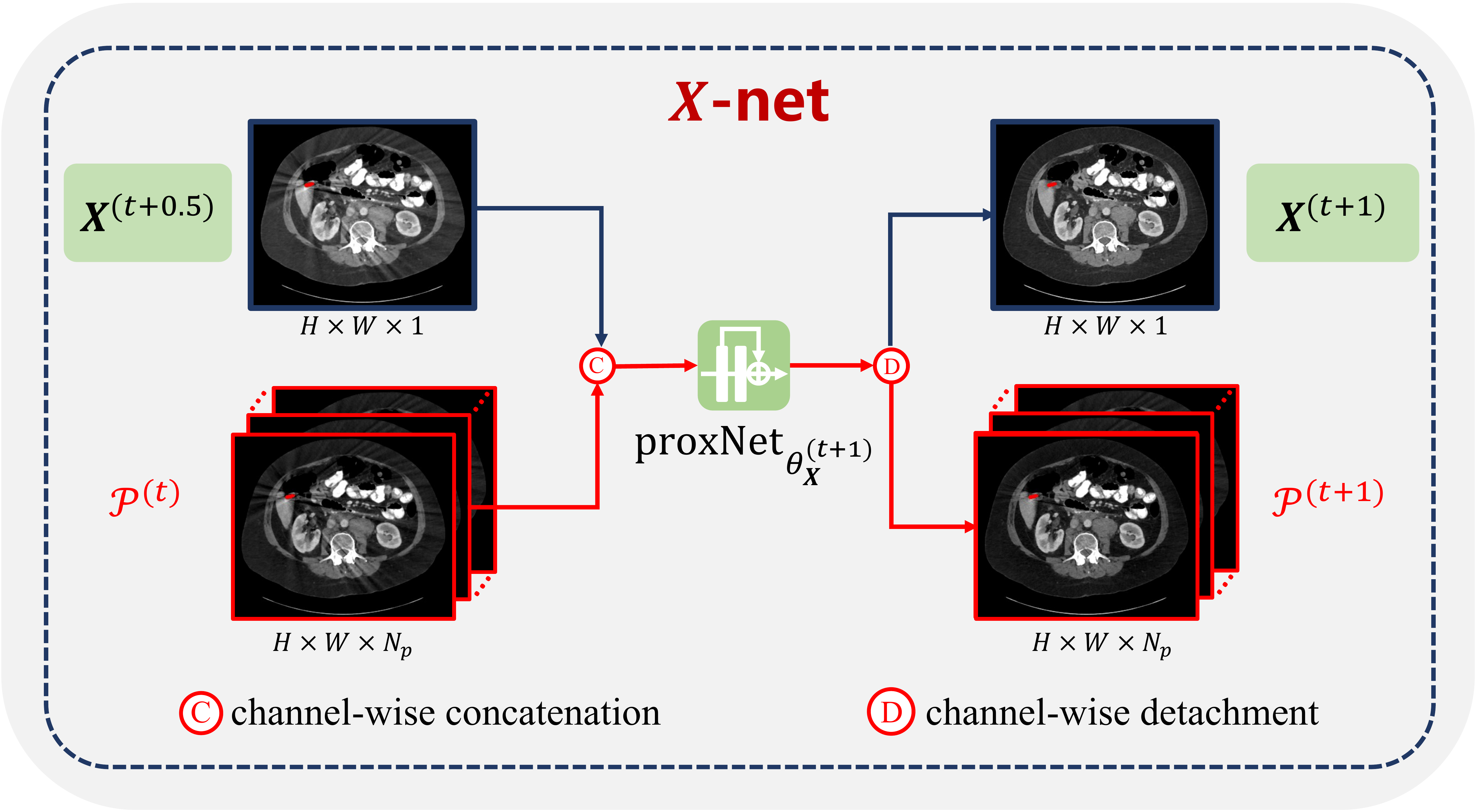}
  \end{center}
  \vspace{-1mm}
     \caption{During the network implementation, this channel expansion operation is implemented for $\X$-net in Fig.~\ref{fignet}.}
  \label{figxnet}
    \vspace{-2mm}
\end{figure}

\subsection{Channel Expansion}
The proposed network framework is shown in Fig.~\ref{fignet} where $\X^{(T)}$ is the final reconstructed CT image.
From the $\X$-net, the original input of the proximal network $\text{proxNet}_{\theta_{\X}^{(t+1)}}(\cdot)$ is $\X^{(t+0.5)}$ with only one gray channel. For the effective feature extraction and the better reconstruction of artifact-reduced CT image, we introduce a channel expansion operation during network implementation, as shown in Fig.~\ref{figxnet}. Specifically, $\mP^{(t)}$ is an auxiliary variable for channel expansion, which is updated together with $\X^{(t+0.5)}$. Hence, the updated input of $\text{proxNet}_{\theta_{\X}^{(t+1)}}(\cdot)$ consists of $(1+N_{p})$ channels (by the channel-wise concatenation). In all our experiments, $N_{p}=32$. Correspondingly, for the output of $\text{proxNet}_{\theta_{\X}^{(t+1)}}(\cdot)$, we execute the channel-wise detachment operation to obtain the updated $\X^{(t+1)}$ and $\mathcal{P}^{(t)}$.

\subsection{Initialization}
From Fig.~\ref{fignet} and Fig.~\ref{figxnet}, to proceed the iterative reconstruction process, we need to initialize $\K^{(0)}$, $\mM^{(0)}$, $\mP^{(0)}$, and $\X^{(0)}$. Since only the metal-corrupted CT image $\Y$ is accessible, we can directly initialize $\mP^{(0)}$ and $\X^{(0)}$ and then utilize them to initialize $\K^{(0)}$ and $\mM^{(0)}$.
% based on the corresponding iterative rules.

\paragraph{Initializing $\mP^{(0)}$ and $\X^{(0)}$.} 
Motivated by the aforementioned channel expansion, we can first execute the channel-wise concatenation between $\X_{LI}$ and $\mC_{p}\otimes\X_{LI}$, and feed the concatenated result to the deep network $\text{proxNet}_{\theta_{\X}^{(0)}}(\cdot)$. Then, with the channel-wise detachment operation, we can easily obtain the $\mP^{(0)}$ and $\X^{(0)}$. Here $\X_{LI}$ is the resorted CT image by the conventional linear interpolation (LI) based method~\cite{kalender1987reduction}; $\mC_{p}$ is a common convolutionl layer where the size of convolutional kernel is $k \times k \times 1 \times N_{p}$ (in our experiment, the value is set to $3 \times 3 \times 1 \times 32$); $\mbox{proxNet}_{\theta_{\X}^{(0)}}(\cdot)$ is a ResNet, with the same structure to $\mbox{proxNet}_{\theta_{\X}^{(t)}}(\cdot)$ as described in Sec. 3 of the main text.

\paragraph{Initializing $\K^{(0)}$ and $\mM^{(0)}$.} 
From Fig.~\ref{fignet}, it is observed that only with $\mP^{(0)}$ and $\X^{(0)}$, we still cannot execute the initialization $\K^{(0)}$ and $\mM^{(0)}$. To solve the problem, we first disentangle $\mM$ and $\K$ for the initialization of $\mM^{(0)}$. Then, with $\mM^{(0)}$, $\mP^{(0)}$, and $\X^{(0)}$, we can easily get the $\K^{(0)}$. 

Specifically, the original optimization problem is
\begin{equation}\label{o1}
\begin{split}
\min_{{\K,\mM,\X}}&\left\|\I\odot\left(\Y\!-\!\X\!-\!\left(\mD\ast\mK\right)\!\otimes\!\mM\right)\right\|_{F}^{2}\\&~~+\alpha f_{1}(\K)\!+\beta f_{2}(\mM)\!+\!\gamma f_{3}(\X)\\
&\hspace{-10mm} \text{subject to}~~~ \left\|\K_{n}\right\|_{2}=1, n =1,2,\ldots, N,
\end{split}
\end{equation}
\normalsize
where $\mD\in\mathbb{R}^{p\times p \times d}$; $\K \in\mathbb{R}^{d \times N}$; and $\mM\in\mathbb{R}^{H\times W\times N}$. In our experiments, $p=9$, $d=32$, and $N=6$.

To disentangle $\mM$ and $\K$, we ignore the weighting mechanism and then simplify the problem ~\eqref{o1} as:
\begin{equation}\label{o2}
\min_{{\mM,\X}}\left\|\I\odot\left(\Y\!-\!\X\!-\!\mD\!\otimes\!\mM\right)\right\|_{F}^{2}+\beta f_{2}(\mM)\!+\!\gamma f_{3}(\X),
\end{equation}
\vspace{1mm}
\normalsize
note that in this simplified case, $d=N=32$.

By adopting the similar solution process in Sec. 3 of the main text and executing the corresponding unfolding operation upon the derived algorithm, we can construct the deep unfolding sub-networks for solving the problem~\eqref{o2}. The network computations are sequentially composed of:
\begin{equation}\label{om1}
 \mM^{(t+1)} = \text{proxNet}_{\theta_{\mM}^{(t+1)}}\left(\mM^{(t+0.5)}\right),   
\end{equation}
\begin{equation}\label{ox1}
\X^{(t+1)} =\mbox{proxNet}_{\theta_{\X}^{(t+1)}}\left(\X^{(t+0.5)}\right),
\end{equation}
where
\begin{equation}
\mM^{(t+0.5)}=\mM^{(t)} \!-\! \eta_{2}\mD\otimes^{T}\!\left(\I\odot\left(\mD \otimes \mM^{(t)}\!+\!\X^{(t)}\!-\!\Y\right)\right),   
\end{equation}
\begin{equation}
    \X^{(t+0.5)}\!=\!\left(\bm{1}\!-\!\eta_3\I\right) \odot\X^{(t)}
  +\eta_{3}\I\odot\left(\!\Y\! \!-\!\!\mD \!\otimes\! \mM^{(t+1)}\!\right).
\end{equation}

Based on Eq.~\eqref{om1} and the aforementioned $\X^{(0)}$, we can get the initial estimation of $\mM^{(0)}$.
Note that, for a better initialization, we further execute an extra iteration sequentially consisting of Eqs.~\eqref{om1}~\eqref{ox1} to refine the $\mM^{(0)}$ and $\X^{(0)}$.

With $\mM^{(0)}$, $\mP^{(0)}$, and $\X^{(0)}$, based on the original computations involved in $\K$-net, \emph{i.e.},
{{$\K^{(t+1)} = \text{proxNet}_{\theta_{\K}^{(t+1)}}\left(\K^{(t+0.5)}\right)$}} (see Sec. 3 of the main text), we can get the initialization of $\K^{(0)}$. Note that in our experiments, the initialization $\mM^{(0)}$ obtained by Eq.~\eqref{om1} for the simplified problem~\eqref{o2} has 32 channels. To execute the regular iterative computation for the original problem ~\eqref{o1}, we simply select the first six channels as $\mM^{(0)}$ for $\mM$-net shown in Fig.~\ref{fignet}.

\section{More Details about Data Simulation}
Following the simulation protocol in~\cite{yu2020deep}, we synthesize the paired metal-free/metal-corrupted CT images by randomly choosing 1,200 clean CT images from the DeepLesion dataset (mainly focusing on abdomen and thorax)~\cite{yan2018deep} and collecting 100 metal masks with diverse types from~\cite{zhang2018convolutional}. 
% Specifically, we separate the 100 masks as three parts, \emph{i.e.}, 80 ones together with the 900 clean CT images for training,  10 ones together with the 100 clean CT images for validation, and 10 ones together with the remaining 200 clean CT images for testing.
During the synthesis of metal-affected CT images, we consider different factors, including poly-chromatic X-ray, partial volume effect, beam hardening, and Poisson noise, which follow the existing studies~\cite{zhang2018convolutional,liao2019adn,lin2019dudonet,yu2020deep,lyu2020dudonet++}. 640 projection views are uniformly spaced between 0-360 degrees. The size of the synthesized CT images is $416\times416$ pixels and the size of the corresponding sinogram data is $641\times640$ pixels.

\begin{minipage}{8.2cm}
\begin{algorithm}[H]
\scriptsize
	\renewcommand{\algorithmicrequire}{\textbf{Input:}}
	\renewcommand{\algorithmicensure}{\textbf{Output:}}
	\caption{\small{Training ACDNet for MAR}}
	\label{alg1}
	\begin{algorithmic}[1]  \small
		% \STATE {\bfseries Input:} Training data $\mathcal{D}$, meta set $\mathcal{D}^{(m)}$, batch size $n,n^{(m)}$, max iterations $T$.
		%   \STATE {\bfseries Output:} Student model parameters $w^{(T)}$
		\REQUIRE  $\K^{(0)}$, $\mM^{(0)}$, $\X^{(0)}$, training data $\left \{\boldsymbol{\Y}_{n},\boldsymbol{\X}_{n},\boldsymbol{\I}_{n}\right \}_{n=1}^{N}$, batch size, patch size, initial learning rate
		\ENSURE Network parameters $\{\theta_{\K}^{(t)},\theta_\mM^{(t)},\theta_{\X}^{(t)}\}_{t=1}^{T}$, stepsizes $\{\eta_{i}\}_{i=1}^{3}$, and the common dictionary $\mD$
		%\REPEAT
		%\STATE Initialize classifier network parameter $\mathbf{w}^{(0)}$ and Meta-Weight-Net parameter $\Theta^{(0)}$.
        \WHILE {The loss in Eq.~(15) of main text is not convergent}
		\FOR{$t=0$ {\bfseries to} $T-1$}
		\STATE $\K$-net:  ~~{{$\K^{(t+1)}= \text{proxNet}_{\theta_{\K}^{(t+1)}}\left(\K^{(t+0.5)}\right)$}}
		\STATE $\mM$-net: ~{{$\mM^{(t+1)} = \text{proxNet}_{\theta_{\mM}^{(t+1)}}\left(\mM^{(t+0.5)}\right)$}}
        \STATE $\X$-net:  ~~$\X^{(t+1)} =\mbox{proxNet}_{\theta_{\X}^{(t+1)}}\left(\X^{(t+0.5)}\right)$
		\ENDFOR
        \STATE Update ACDNet (\emph{i.e.}, $T$-stage $\K$-net, $\mM$-net, and $\X$-net).
        \ENDWHILE
	\end{algorithmic}
\end{algorithm}
\end{minipage}
\normalsize

\section{More Implementation Details}
The training process is listed as Algorithm 1. We execute two testing tasks on CLINIC-metal, including metal artifact reduction (MAR) task and the downstream pelvic fracture segmentation task.  For the segmentation task, we first need to train a U-Net on clean pelvic data. The training details are described as:

A publicly available clinical metal-free dataset--CLINIC~\cite{liu2020deep}, with the annotations of multi-bone structures, is adopted. It consists of 103 clean volumes (35,518 slices). The Adam optimizer~\cite{Kingma2014Adam} is adopted for network optimization. The initial learning rate is $2\times10^{-4}$ and divided by 2 every 20 epochs. The network converges after 42 epochs of training with the batch size of 12. The total loss function is the combination of cross-entropy loss and dice loss.

% We randomly divide them into two sets (\emph{i.e.}, training set and testing set) according to the ratio of 70:30. 

\begin{figure*}[!t]
  \begin{center}
  %\vspace{-3mm}
     \includegraphics[width=1\linewidth]{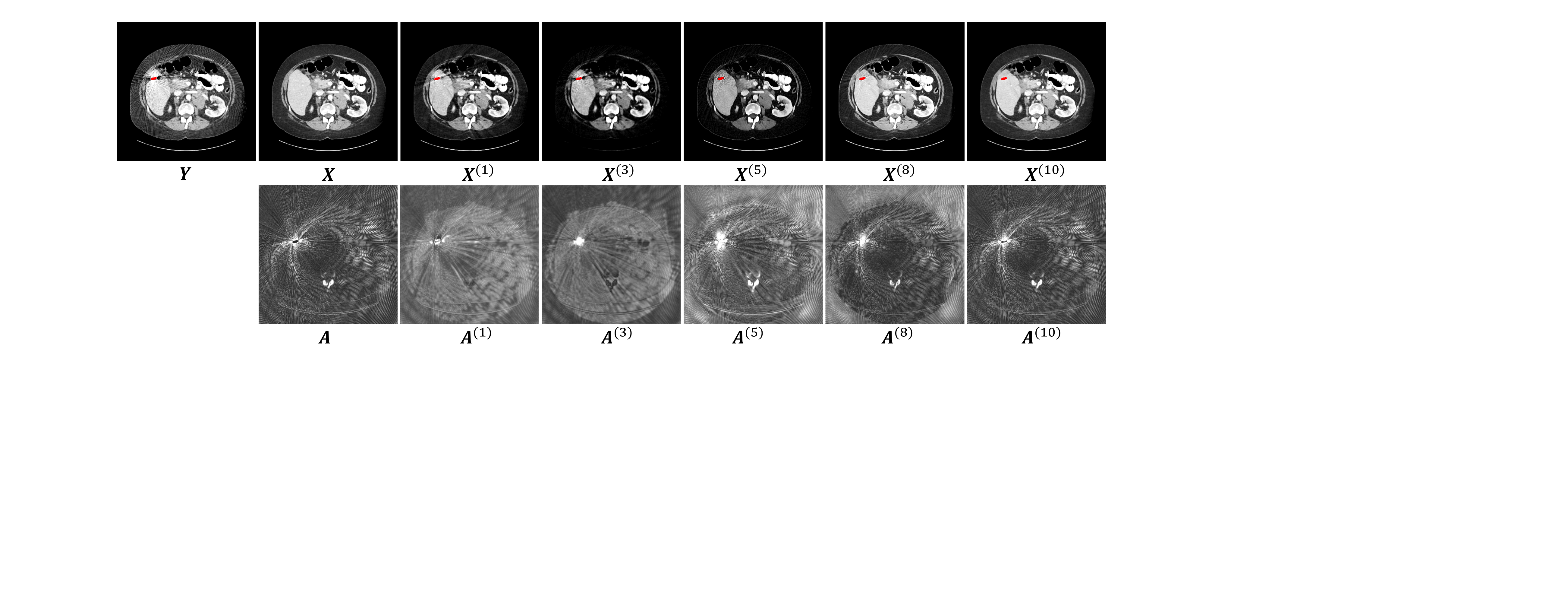}
  \end{center}
  \vspace{-4mm}
     \caption{Visualization of artifact-reduced image $\X^{(t)}$ and artifact $\A^{(t)}$ extracted by our ACDNet at the stage $t$. The total stage $T$ is 10 and $\A$ is the groudtruth artifact, expressed as $\left(\Y-\X\right)\odot\I$.} % (Training-testing domain match)
  \label{figmodel}
  \vspace{-0mm}
\end{figure*}

\section{More Analysis about ACDNet}
\subsection{Interpretability Verification}
To verify the specific interpretability of our proposed ACDNet for the MAR task, we visualize the learning process, as shown in Fig.~\ref{figmodel}. It is easily observed that with the increase of iterative stage $t$, the extracted artifact layer $\A^{(t)}$ is approaching the ground truth artrifact $\A$ and the reconstructed CT image $\X^{(t)}$ is approaching the ground truth clean image $\X$.  This results finely substantiate that the mutual promotion learning of $\K$-net, $\M$-net, and $\X$-net and the guidance of the explicit weighted prior model make the entire network framework optimize in a right direction for accurately identifying metal artifacts. Such interpretable learning process is the intrinsic characteristic of our iterative optimization framework. Actually, compared to heuristic network designs, our ACDNet is more transparent and interpretable since every network module is accordingly built based on the optimization algorithm.

% \subsection{Performance of Convergence Process}
% \begin{figure}[t]
%  \begin{center}
%     \includegraphics[width=1\linewidth]{PIC/rebutttal_conver.pdf}
%  \end{center}
%  \vspace{-3mm}
%     \caption{Average PSNR (dB) during 200 training epochs.}
%  \label{figtrain}
%     \vspace{-2mm}
% \end{figure}
% Besides, we visualize the performance of convergence process during the first 200 training epochs. As shown in Fig.~\ref{figtrain}, compared to the latest SOTA MAR methods, our proposed method can more steadily converge to better performance. 
% This can be mainly attributed to the sound guidance of the derived algorithm and that our method is only based on CT images without complicated Radon transformation between projection data and CT image.

\begin{figure*}[!t]
  \begin{center}
  %\vspace{-3mm}
     \includegraphics[width=1\linewidth]{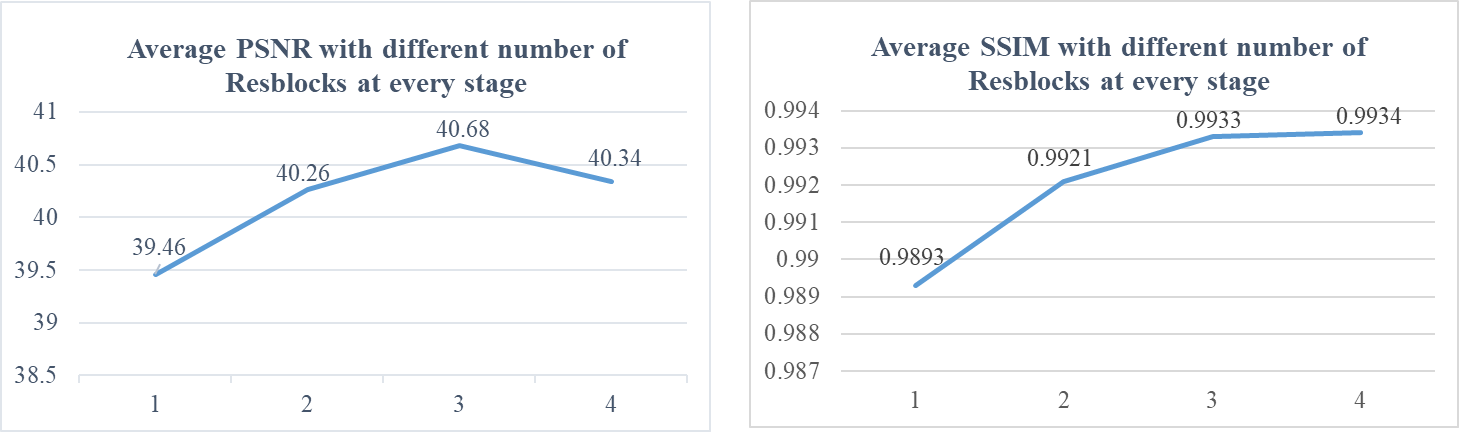}
  \end{center}
  \vspace{-2mm}
     \caption{The average PSNR/SSIM of our method on the synthesized DeepLesion dataset with the different numbers of Resblocks involved in deep proximal networks $\text{proxNet}_{\theta_{\X}^{(t)}}(\cdot)$.} % (Training-testing domain match)
  \label{figabla}
  \vspace{-1mm}
\end{figure*}

\subsection{Ablation Study}
\paragraph{The number of Resblocks.}
Fig.~\ref{figabla} shows the variation on MAR performance of our framework with the different numbers of Resblocks in deep proximal networks $\text{proxNet}_{\theta_{\X}^{(t)}}(\cdot)$ at every stage, where the total number $T$ of iterative stages is 10. Note that $\text{proxNet}_{\theta_{\mM}^{(t)}}(\cdot)$ adopts the same number of Resblocks to $\text{proxNet}_{\theta_{\X}^{(t)}}(\cdot)$. Based on the results, we adopt 3 Resblocks to construct the backbone of deep proximal networks as described in Sec. 3 of the main text. Note that we only adopt 1 Resblock for building $\text{proxNet}_{\theta_{\K}^{(t)}}(\cdot)$ to reduce the network parameters of the whole framework.

\paragraph{The Updating Order.}
% By changing the orders of $\K$-net, $\mM$-net, and $\X$-net, the average PSNRs on synthesized DeepLesion are: ($\K$-net, $\mM$-net, $\X$-net,  40.75dB); ($\mM$-net, $\K$-net, $\X$-net, 40.42dB); ($\K$-net, $\X$-net, $\mM$-net, 40.54dB); ($\mM$, $\X$-net, $\K$-net, 40.35dB); ($\X$-net, $\K$-net, $\mM$-net, 40.24dB); ($\X$-net, $\mM$-net, $\K$-net, 40.67dB). 
\begin{table}[t]
\centering
    \begin{tabular}{c|c}
    \hline
     Optimization order      & PSNR (dB) \\
    \hline
    $\K$-net,  $\mM$-net, $\X$-net  &40.68\\
    $\mM$-net, $\K$-net,  $\X$-net  &40.42\\
    $\K$-net,  $\X$-net,  $\mM$-net &40.54\\
    $\mM$-net, $\X$-net,  $\K$-net  &40.35\\
    $\X$-net,  $\K$-net,  $\mM$-net &40.24\\
    $\X$-net,  $\mM$-net, $\K$-net  &40.64\\
    \hline
    \end{tabular}
     \caption{Average PSNR on synthesized DeepLesion under different updating orders.}
\label{taborder}
\vspace{-2mm}
\end{table}
\normalsize
Table~\ref{taborder} lists the performance of our proposed ACDNet with different updating orders. As seen, our method is insensitive to the order. 

\section{More Experiments}
\subsection{Performance Comparisons}

\begin{figure*}[t]
  \begin{center}
  %\vspace{-3mm}
     \includegraphics[width=1\linewidth]{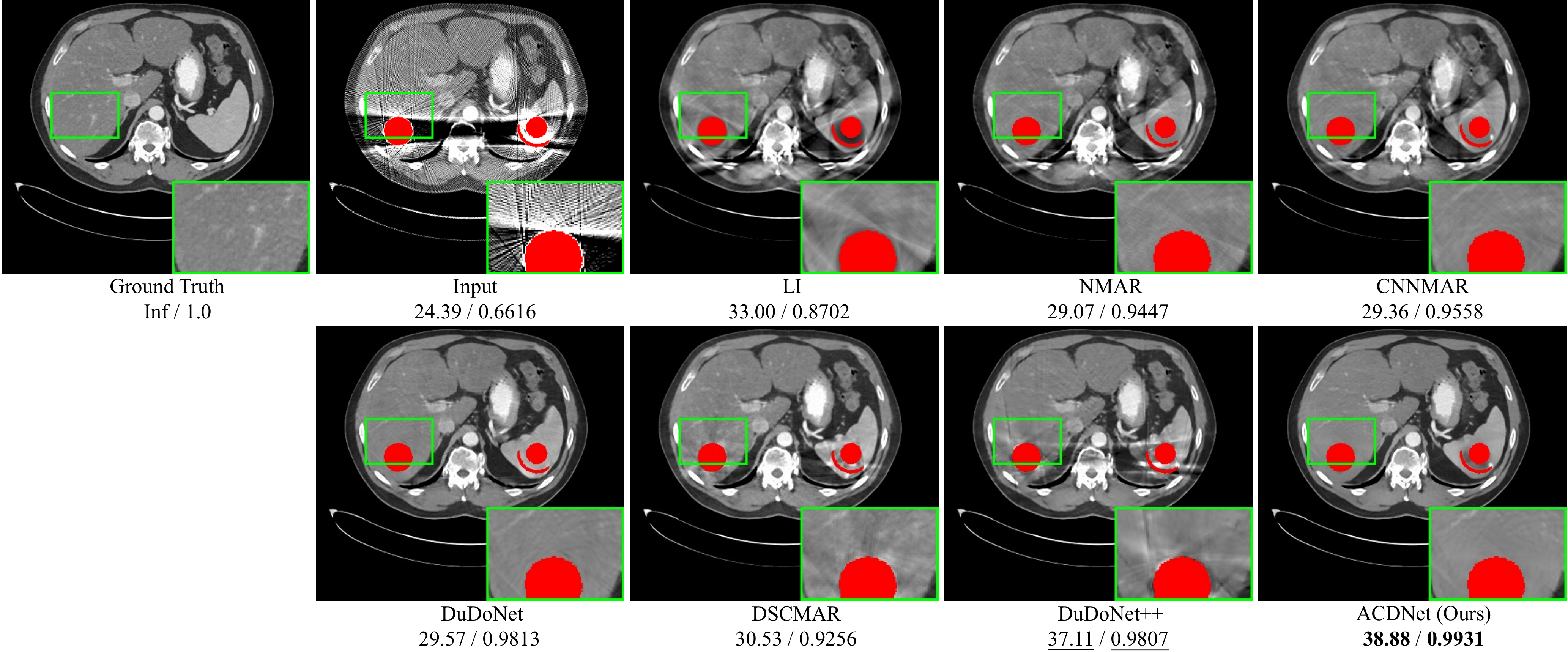}
  \end{center}
  \vspace{-3mm}
     \caption{Performance comparison of different MAR approaches on a metal-corrupted CT image with large metals selected from the synthesized DeepLesion data. PSNR (dB)/SSIM below is listed for reference. The red pixels stand for metallic implants.} % (Training-testing domain match)
  \label{figsynbig}
  \vspace{-0mm}
\end{figure*}

\begin{figure*}[!t]
  \begin{center}
  %\vspace{-3mm}
     \includegraphics[width=1\linewidth]{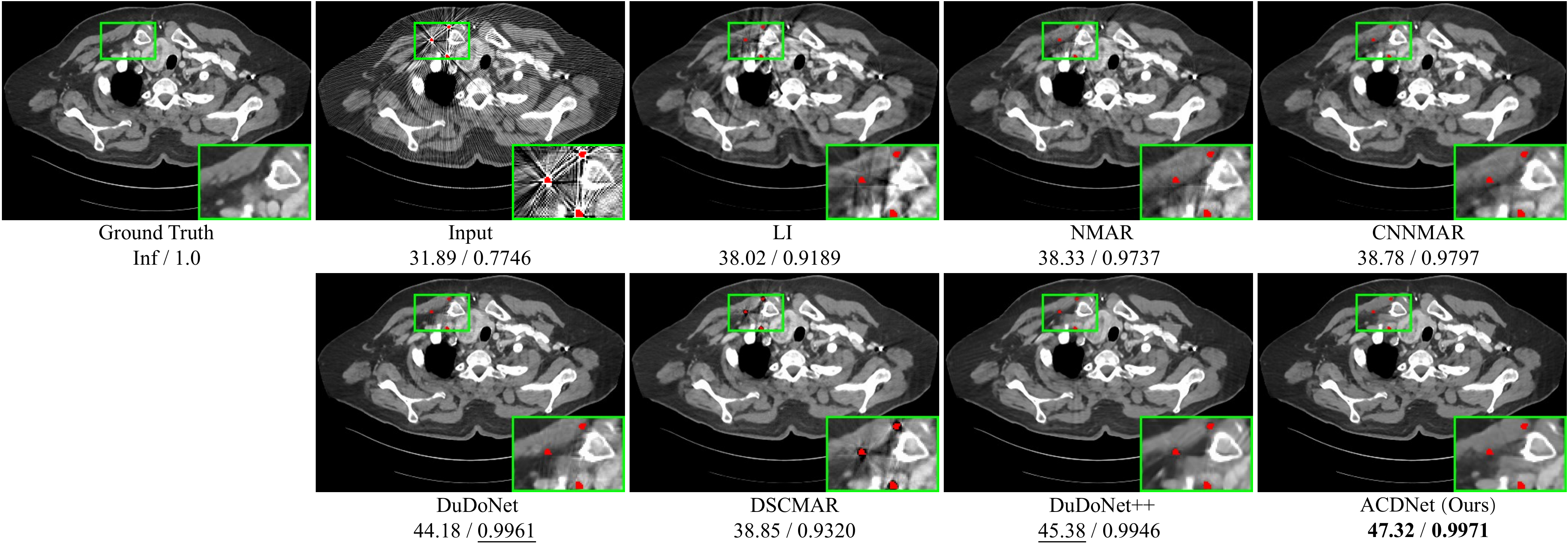}
  \end{center}
  \vspace{-3mm}
     \caption{Performance comparison of different MAR approaches on a metal-corrupted CT image with small metals selected from the synthesized DeepLesion data. PSNR (dB)/SSIM below is listed for reference.} % (Training-testing domain match)
  \label{figsynsmall}
  \vspace{1mm}
\end{figure*}

\paragraph{DeepLesion Data.}
We provide more visual results of different MAR approaches on different metallic implants with varying sizes as shown in Fig.~\ref{figsynbig} and Fig.~\ref{figsynsmall}. It is clear that, among the comparing methods, the proposed framework achieves the better reconstruction results.

\paragraph{Dental Data.}
Table~\ref{tabdental} reports the quantitative performance of different MAR methods on the synthesized dental CT images shown in Fig. 4 of main text. As seen, it shows the excellent cross-body-site generalizability of our method.
\begin{table*}[!htp]
\centering
\small
\setlength{\tabcolsep}{5.2pt}
    \begin{tabular}{c|c|c|c|c|c|c|c|c}
    \hline
    Figure        & Input & LI            &NMAR    &CNNMAR &DuDoNet & DSCMAR  & DuDoNet++ & ACDNet (Ours)    \\
    \hline
    (a)   & 34.52/0.8839  & 31.14/0.8783  & 31.69/0.9003   & 35.23/0.9532 & 35.67/0.9685  & \underline{36.80}/0.9709  & 36.04/\underline{0.9857}  & \textbf{41.41}/\textbf{0.9903} \\
    (b) & 36.46/0.9282 & 33.93/0.9435 & 34.86/0.9626 & 36.37/0.9761 & 39.38/0.9744 & 37.11/0.9806 & \underline{40.71}/\underline{0.9903} & \textbf{45.40}/\textbf{0.9954}\\
    \hline
    \end{tabular}
    \caption{PSNR (dB)/SSIM of different MAR methods on the synthesized dental CT images shown in Fig.4 of main text.}
\label{tabdental}
\vspace{-1mm}
\end{table*}
\normalsize

\begin{table*}[!t]
%   \vspace{-1mm}
\centering
\setlength{\tabcolsep}{9pt}
% \small
\begin{tabular}{l|cccccccc}
\hline
Bone       & Input   & LI    & NMAR  & CNNMAR & DuDoNet& DSCMAR & DuDoNet++  & Ours\\
\hline
Sacrum &92.47  & 90.86   & 91.51 & 92.44                     &93.26  &92.52  & \textit{\textbf{93.50}}  &\textbf{94.34} \\
Left hip     & 95.43  &93.91 &94.27                   &94.85                     &96.11  &95.33  &\textit{\textbf{96.17}}     &\textbf{96.75}   \\
Right hip    & 87.47  &91.23 &91.68                  &92.50                     & \textit{\textbf{93.89}} &93.22  & 93.79     &\textbf{94.98}  \\
Lumbar spine & 94.43 &94.53 &94.64                   &94.89                     &95.51  &94.75  &\textit{\textbf{95.64}}      &\textbf{95.77} \\
\hline
Average DC &92.45  &92.63 &93.03                   &93.67                     &94.69  &93.96  &\textit{\textbf{94.77}}  &\textbf{95.46} \\
\hline
\end{tabular}
\caption{The average Dice coefficient (DC) (\%) results on the clinical downstream segmentation task.}
\label{tabclinic}
    \vspace{-2mm}
\end{table*}
\paragraph{Clinical Data.}
Fig.~\ref{figclinic} presents the visual comparison of MAR performance on a clinical image selected from CLINIC-metal, where deep MAR methods are trained on synthesized DeepLesion data.

Table~\ref{tabclinic} reports the average segmentation accuracy on 14 artifact-removed CLINIC-metal volumes. Please note that this segmentation task is very challenging due to the
large gap between DeepLesion and clinical pelvic CT. Then
the quantitative improvement for a new method is generally
not that significant, like the gain (0.08) of DuDoNet++
over DuDoNet. Thus, the looking-like “small” improvement
(0.69) of our method over DuDoNet++ is rational to support
its effectiveness. Actually, visual comparison should be more
intuitive. Fig.~\ref{figseg} displays the downstream segmentation results of an artifact-reduced CLINIC-metal volume, which are reconstructed by different MAR approaches. This more intuitively shows clinical benefits of our method.

% The thresholding method is widely adopted in the current SOTA methods.
% Nevertheless, our method is observed to be robust to metal segmentation. Our method uses only CT image and there is no necessary connection between metal segmentation and artifact modelling; 2) Our method consistently achieves better performance under different types of metals. Smaller thresholding possibly causes that more tissues are mistakenly regarded as metal regions and then the reconstructed images would lose mass content. Analysis of the thresholding method is beyond our focus and we will do specific exploration in the future.

% \noindent\textbf{Q5.} The soundness is poor and the reproducibility is fair.

% \noindent\textbf{A5.}  The reproducibility is friendly. First, the proposed unfolding network is rationally built based on the optimization algorithm. Such unfolding techniques have achieved great success in many vision tasks, such as [35][43], which have attracted much attention with released codes. Besides, the training protocol described in Sec.~3.2 of SM is maintained for all the experiments implemented on synthetic data and our model can always achieve good performance, which shows fine robustness of our method. Moreover, we promise that we will release the source code and the pretrained models once this paper is accepted for publication.

\begin{figure*}[!t]
  \begin{center}
  %\vspace{-3mm}
     \includegraphics[width=1\linewidth]{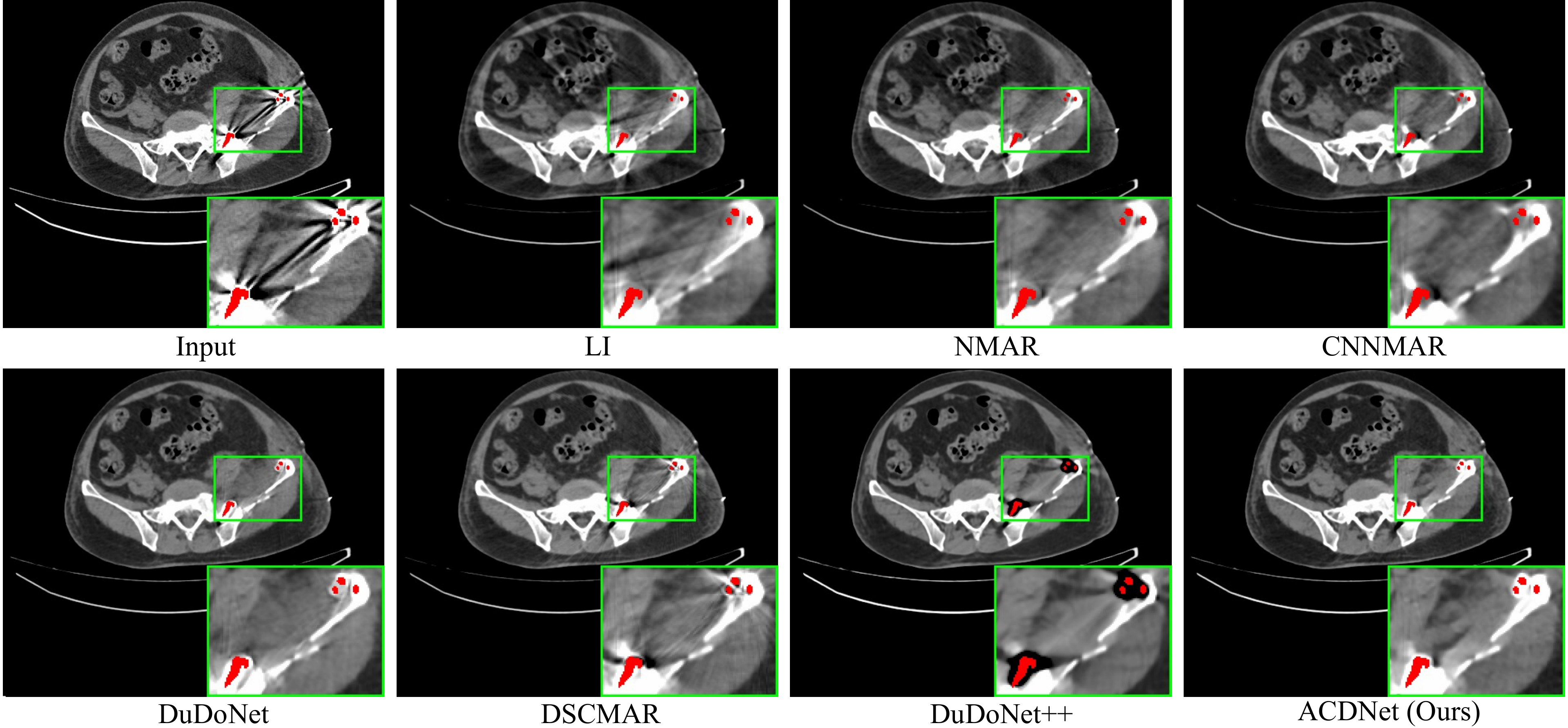}
  \end{center}
  \vspace{-2mm}
     \caption{Generalization results. Performance comparison on a real clinical metal-affected CT image from CLINIC-metal, where deep MAR methods are trained on the synthesized DeepLesion data. The red pixels stand for metallic implants.}
  \label{figclinic}
\end{figure*}
\begin{figure*}[!t]
  \begin{center}
  %\vspace{-3mm}
     \includegraphics[width=1\linewidth]{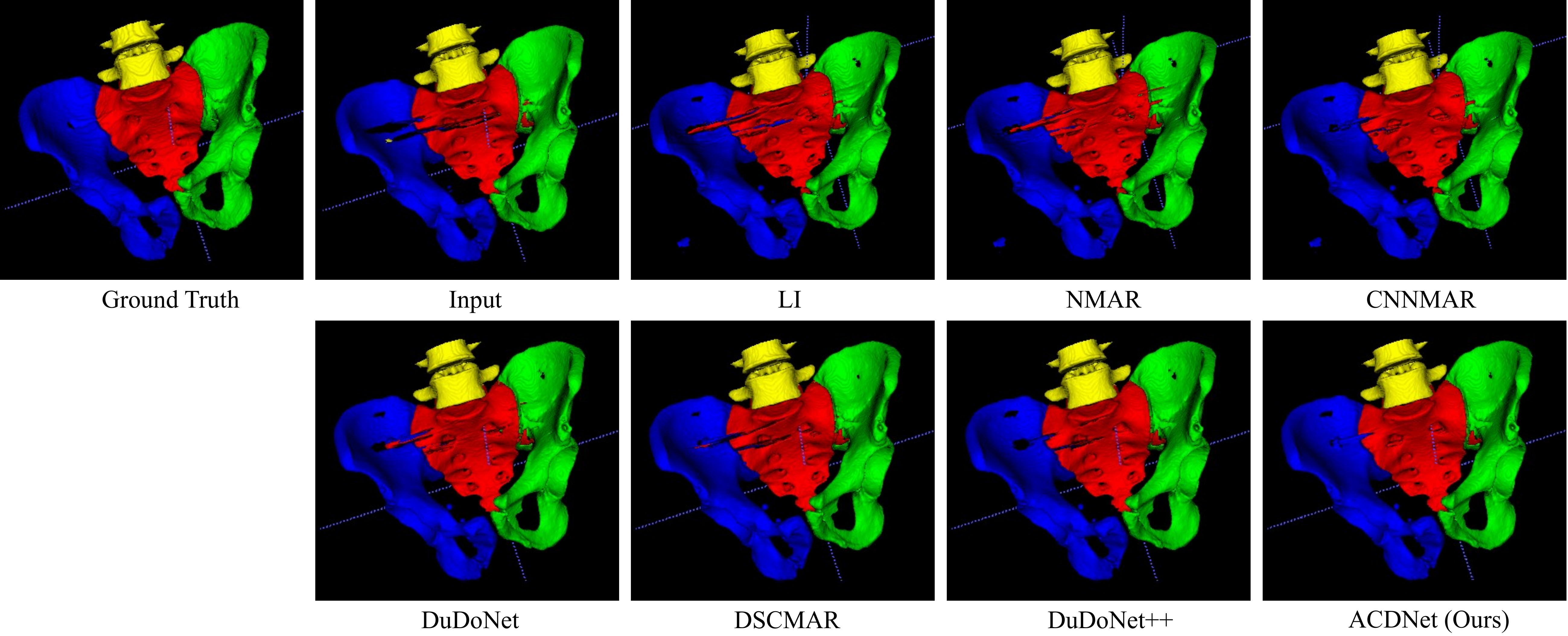}
  \end{center}
  \vspace{-2mm}
     \caption{Downstream segmentation results of an artifact-reduced CLINIC-metal volume, which are reconstructed by different MAR approaches. Note that the deep MAR methods are trained on the synthesized DeepLesion data.} % (Training-testing domain match)
  \label{figseg}
  \vspace{-0mm}
\end{figure*}

\begin{table}[t]
\centering
\small
\setlength{\tabcolsep}{2pt}
    \begin{tabular}{l|c|c|c|c|c}
    \hline
    Table       & CNNMAR &DuDoNet & DSCMAR & DuDoNet++  &Ours  \\
    \hline
    Table 1$^*$  &$<0.001$ & $<0.001$ &$<0.001$   &$<0.001$ & n.a\\
    Table 2  &0.0421& 0.0147 & 0.0391 & 0.0362 & n.a\\
    Table 3 & $<0.001$ &$<0.001$ &$<0.001$ & 0.0203 & n.a\\
    \hline
    \end{tabular}
      \caption{P-values comparison. Note that Table 1 is the one in the main text.}
\label{tabpvalue}
\vspace{-4mm}
\end{table}
\normalsize

\paragraph{P-values Analysis.}
With the paired t-test, the P-values for the quantitative comparisons listed in Table 1 of main text, Table~\ref{tabdental}, and Table~\ref{tabclinic} are reported in Table~\ref{tabpvalue}. It is easily observed that all P-values are less than the significance level 0.05. Thus, our method performs significantly better.

\end{document}